\documentclass[namedreferences]{SolarPhysics}
\usepackage[optionalrh,solaenum,natbib]{spr-sola-addons} 
\usepackage{graphicx,times,url,color}

\newcommand{\aap}{    {\it Astron. Astrophys.}}

\newcommand{\apj}{    {\it Astrophys. J.}}
\newcommand{\apjl}{    {\it Astrophys. J. Lett.}}

\newcommand{\mnras}{  {\it Mon. Not. Roy. Astron. Soc.}}
\newcommand{\nat}{    {\it Nature}}

\newcommand{\solphys}{{\it Solar Phys.}}

\begin{document}
\begin{article}
\begin{opening}

\title{Local Helioseismology of Sunspots: Current Status and Perspectives\footnote{
Invited Review.}}
\author{
        Alexander~G.~Kosovichev} \runningauthor{A.~G.~Kosovichev}
        \runningtitle{Local Helioseismology of Sunspots}
        \institute{W.W.Hansen Experimental Physics Laboratory,
        Stanford University, Stanford, CA 94305-4085, USA;
           email: \url{sasha@sun.stanford.edu} \\
          }

\begin{abstract}
Mechanisms of the formation and stability of sunspots are among the longest-standing and intriguing puzzles of solar physics and astrophysics. Sunspots are controlled by subsurface dynamics hidden from direct observations. Recently, substantial progress in our understanding of the physics of the turbulent magnetized plasma in strong-field regions has been made by using numerical simulations and local helioseismology. Both the simulations and helioseismic measurements are extremely challenging, but it becomes clear that the key to understanding the enigma of sunspots is a synergy between models and observations. Recent observations and radiative MHD numerical models have provided a convincing explanation to the Evershed flows in sunspot penumbrae. Also, they lead to the understanding of sunspots as self-organized magnetic structures in the turbulent plasma of the upper convection zone, which are maintained by a large-scale dynamics. Local helioseismic diagnostics of sunspots still have many uncertainties, some of which are discussed in this review. However, there have been significant achievements in resolving these uncertainties, verifying the basic results by new high-resolution observations,
 testing the helioseismic techniques by numerical simulations, and comparing results obtained by different methods. For instance, a recent analysis of helioseismology data from the {\it Hinode} space mission has successfully resolved several uncertainties and concerns (such as the inclined-field and phase-speed filtering effects) that might affect the inferences of the subsurface wave-speed structure of sunspots and the flow pattern. It becomes clear that for the understanding of the phenomenon of sunspots it is important to further improve the helioseismology methods and investigate the whole life cycle of active regions, from magnetic-flux emergence to dissipation. The {\it Solar Dynamics Observatory} mission has started to provide data for such investigations.
\end{abstract}


\end{opening}

\section{Introduction}
One of the primary goals of local-area helioseismology is to
investigate subsurface structures and dynamics of the quiet-Sun and
active regions. Of particular interest is the investigation of the
subsurface structures and flows beneath sunspots. The main goal of
these studies is to understand the mechanism of formation and stability
of sunspots. From the physical point of view, sunspots represent stable,
self-organized, magnetic structures in the turbulent convective plasma.
Such magnetic self-organization phenomena are of significant
interest in physics and astrophysics. In addition, sunspot regions
are the primary source of solar disturbances and energetic events.

Previously, the structure of sunspots was studied only from
observations of the solar surface. Local helioseismic
techniques have provided measurements of variations of travel times and
oscillation frequencies associated with the subsurface structure and
dynamics of sunspots. These measurements open an opportunity for
inferring the subsurface properties of sunspots by inversion of
the travel times and frequency shifts. Initially, the local helioseismology studies of
sunspots were developed without much theoretical support, using
physical intuition and simple models of wave propagation. The
criticism of these studies was also based on highly simplified
models and arguments. But, recently, substantial progress has been
made in developing realistic MHD simulations of dynamics of the
turbulent magnetized plasma. Creating a synergy of the local-helioseismology
measurements and the simulations is the most recent
development in this field, which undoubtedly will lead to new understanding of the sunspot phenomenon.

The inversion (tomographic) procedures of seismology and
helioseismology are well-developed, particularly, when the inverse
problem is reduced to the solution of integral equations relating
 variations  of the oscillation frequencies and travel times
to perturbations of  interior properties. However, in the
case of sunspots these relationships have not been well-established, and
thus the interpretation of helioseismic measurements and inversion
results remains uncertain.

The main reasons for the helioseismic uncertainties arise from a complex
interaction of solar oscillations with strong magnetic fields of
sunspots, non-uniform distribution of wave sources in the sunspot areas,
and uncertainties of the helioseismic measurement procedures. The initial
inferences have been made by using relatively simple physical
relations derived from a ray-path approximation or a first Born
approximation. These results caused a significant interest and
discussions. With the rapid progress of supercomputing, an important
role is being played by direct MHD numerical simulations, which
provide opportunities for understanding the physics of sunspots,
wave excitation and propagation in sunspot regions, and also provide artificial data for testing
the local-helioseismology inferences.

Currently, the important work
of developing the  synergy between the numerical simulations and local
helioseismology measurements and inversions is still in an initial stage,
but first important results have been obtained. This review briefly
describes the current status of this effort and discusses some key
problems of the local helioseismic diagnostics of sunspots.

\section{Models of Magnetic Structures and Sunspots}

\begin{figure}
\centerline{\includegraphics[width=1.0\textwidth]{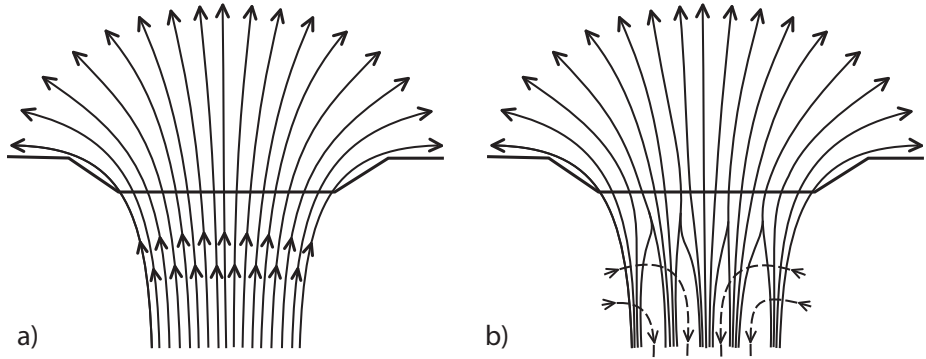}
} \caption{Illustrations of (a) monolithic and (b) cluster sunspot
models proposed by \citet{Parker1979}. Solid curves show magnetic field lines;
dashed curves show subsurface, which provide accumulation of
magnetic flux and stability of sunspots.} \label{fig1} \end{figure}

The mechanism of formation of sunspots is not yet established.
However, the dynamical nature of sunspots is apparent. It has been realized long ago that sunspots are a product of complex interactions of turbulent convection, radiation,
and magnetic field.
\citet{Cowling1946} analyzing Mt.Wilson observations of the growth
and decay of sun\-spots first suggested that the sunspots structure
cannot be magnetostatic, and that the sunspot magnetic field lines
are "bunched together into the spot as a results of mass motion".
\citet{Parker1979} noticed that the formation of sunspots occurs due
to coalescence of magnetic elements in regions of magnetic-flux
emergence, and that the sunspot growth continues as long as "new
magnetic flux (in the form of the small individual flux tubes)
continues to emerge through the surface of the Sun." He argued that
the subsurface structure of sunspots most likely represents loose
bundles of magnetic flux tubes confined by converging downflows in
the surrounding plasma beneath the solar surface (Figure~\ref{fig1}).

In additions, theoretical investigations of magnetostatic models of
sunspots \citep{Meyer1977,Jahn1992,Moreno-Insertis1989} showed that
such models are intrinsically unstable, and thus dynamical effects
are required for the stability of sunspots. However, theoretical
modeling of the sunspot structure and dynamics, particularly of the
cluster type, is a very difficult task. Therefore, magnetostatic
(monolithic) models have been constructed for the purpose of
comparison with spectropolarimetric observations
\citep[\textit{e.g.}][]{Maltby1986}, and more recently, for testing local
helioseismology techniques
\citep{Khomenko2009,Parchevsky2010,Cameron2010}. These models play an
important role for the understanding of the
wave interaction with magnetic field and for  helioseismology
testing. However, it is important to remember that mass motions and
the filamentary structure of magnetic fields are critical for the
physics of both sunspots and waves propagating in sunspot regions.
The magnetostatic or MHD models, in which the magnetic field lines are held together
by external artificial forces ({\it e.g.} by setting up a boundary
condition at the bottom boundary, which holds the magnetic-field
concentration), cannot be considered as physically complete.
Sunspots represent a self-organized dynamic phenomenon in
turbulent magnetized plasma of the solar convection zone, and this is
very difficult to model.

It is well known that long-lived sunspots develop penumbrae
consisting of almost horizontal filamentary magnetic structures.
Observations clearly show mean outflow in the penumbrae, the
\citet{Evershed1909} effect, and also a moat flow in the surrounding
plasma. At first sight, these flows seem to be consistent with a
diverging circulation flow pattern beneath sunspots. However, the
situation may not be that simple.

\begin{figure}
\centerline{\includegraphics[width=0.7\textwidth]{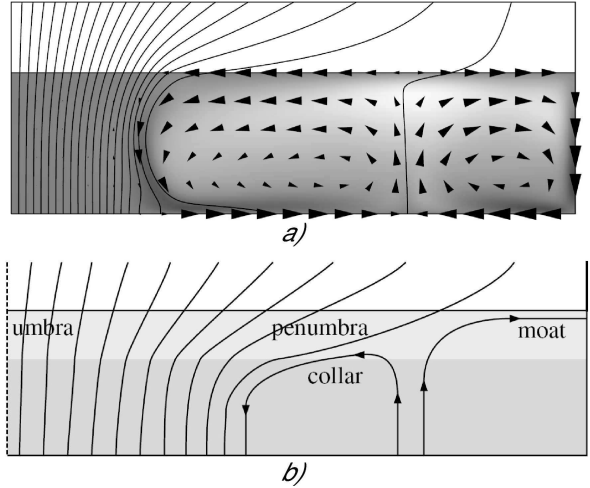}}
 \caption{a) Flow pattern (arrows) and magnetic field lines of an
axisymmetric MHD model of sunspots; b)
 collar and moat flows in this model (after \citet{Hurlburt2000}).}
\label{fig2} \end{figure}

\begin{figure}
\centerline{\includegraphics[width=0.7\textwidth]{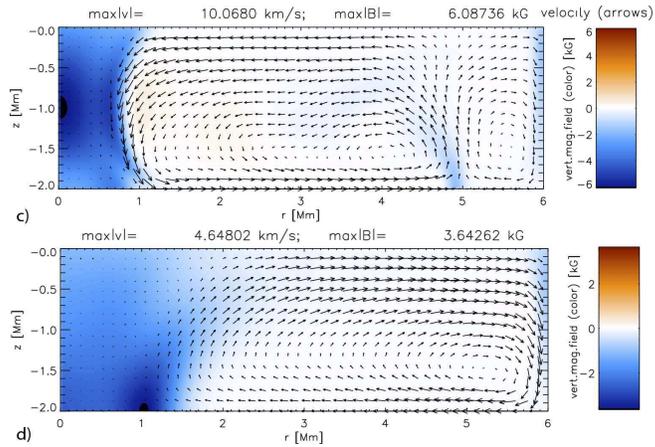}}
 \caption{a) Simulations of axisymmetric sunspot-type structure for solar conditions with subsurface converging flow pattern; b) simulation of a rotating sunspot structure with diverging flows \citep{Hartlep2010}.}
\label{fig3} \end{figure}

In a series of papers,
\citet{Hurlburt2000}, \citet{Botha2006}, \citet{Botha2007},
and \citet{Botha2008} developed
numerical MHD models of the formation of magnetic structures in a
convective layer and concluded that stable magnetic structures can
be formed only by converging flows, and diverging flows inevitably
tore the structures apart. They suggested that the Evershed and moat
flows are confined in a near-surface layer, and that beneath these
there is a converging ``collar'' flow that provides the stability of
sunspots (Figure~\ref{fig2}). The simulations of \citet{Hartlep2010} showed the formation of stable sunspot-like structures is accompanied by strong converging flows. The diverging flows may appear in the case of rotating sunspot structures (Figure~\ref{fig3}), but the converging `collar' flow seems to be essential for sunspot stability.
These simulations were carried for axisymmetrical
configurations. But, recently, \citet{Hurlburt2008} confirmed in 3D
simulations that large-scale magnetic structures can be formed by
inflows driven by a surface cooling. However, these simulations did
not reproduce the sunspot's penumbra. Also, these simulations did not
include the near-surface turbulent convection, which, in general, tends to destroy magnetic configurations.

The physics of sunspot formation, stability, and decay in the
turbulent radiating plasma, and the wave excitation and propagation
through this medium is very complicated. It seems that our best hope
for understanding sunspots is in developing 3D MHD simulations,
which take into account all essential elementary physical processes
including radiative and turbulent effects. With the fast massive
parallel supercomputers becoming more and more available, the solar
MHD simulations are making rapid  progress.

\begin{figure}
\centerline{\includegraphics[width=0.85\textwidth]{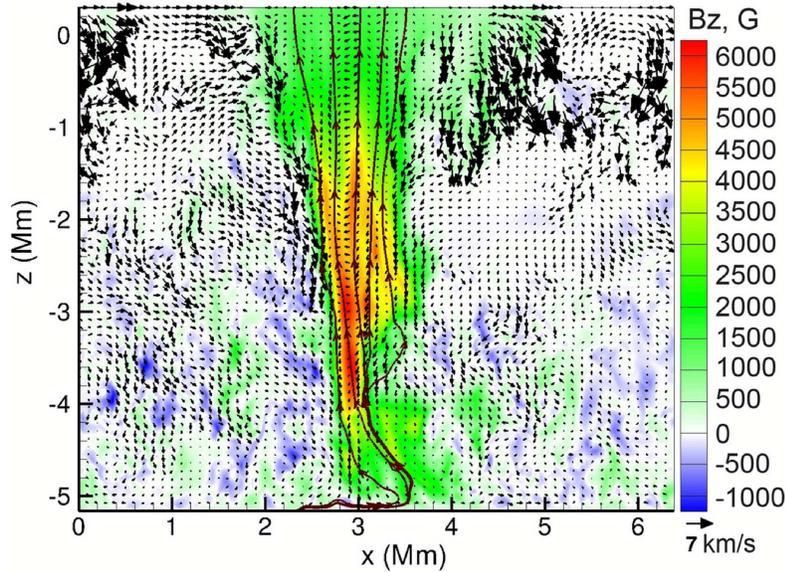}
} \caption{A vertical cut of the 3D MHD simulations, showing the
mass flows (arrows) and the vertical magnetic field strength
(background color) of a stable pore-like structure, spontaneously
formed from an initially vertical magnetic field
\citep{Kitiashvili2010a}. } \label{fig4} \end{figure}

Realistic 3D MHD numerical simulations have been able to reproduce
formation of relatively small pore-like  magnetic structures
in the turbulent upper convective layer
\citep{Stein2003,Vogler2005,Kitiashvili2010}.
The simulations have shown that the concentration of
magnetic field is accompanied by surface cooling and strong
downdrafts and inflows around the magnetic elements (Figure~\ref{fig4}).
This is consistent with the early ideas of
\citet{Schmidt1968}, \citet{Ponomarenko1972}, and \citet{Parker1979}. Such converging
flow pattern
is also established in observations of solar pores, which do not
have penumbra \citep{Sankarasubramanian2003,VargasDominguez2010}.

\begin{figure}
\centerline{\includegraphics[width=0.85\textwidth]{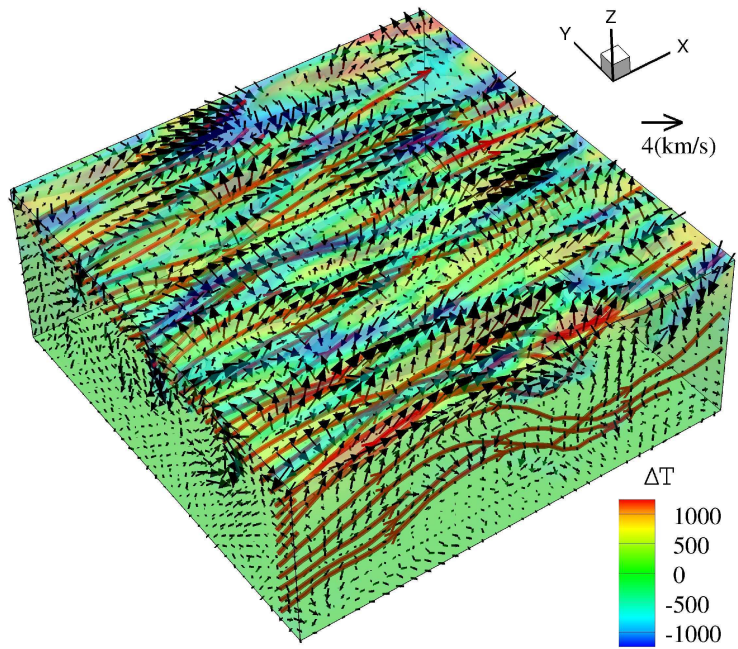}
} \caption{Numerical simulations of magnetoconvection in an inclined
magnetic field, illustrating the origin of the Evershed flow and
the filamentary structure of a sunspot penumbra. The magnetic
field strength is 1000~G; the mean inclination angle is $85^\circ$
from axis $Z$ in the $X-Z$-plane. Arrows show 3D flow velocity;
semi-transparent color background show the variations of
temperature; red curves show magnetic-field lines. The horizontal
size of the box is 6~Mm; the depth is 2~Mm (only the upper part of
the 6-Mm deep simulation domain is shown) \citep{Kitiashvili2009}. }
\label{fig5} \end{figure}

The realistic MHD simulations
\citep{Heinemann2007,Scharmer2008,Rempel2009,Kitiashvili2009,Scharmer2009} have also
led to new understanding of the sunspot penumbra
structure and dynamics.
The simulations convincingly showed that the filamentary structure
and the plasma outflow are a natural consequence of magnetoconvection in
the regions of strong inclined magnetic field (Figure~\ref{fig5}).
Magnetoconvection in the inclined field has properties of waves
traveling in the direction of the field inclination \citep{Hurlburt2000}.
The simulations of \citet{Kitiashvili2009} show that this effect contributes
to the the generation of the organized radial outflow in sunspot penumbrae.
According to the realistic simulations the penumbra outflow (Evershed effect)
represents the overturning convective motions
along the magnetic field, which are also amplified and organized by the traveling
magnetoconvection waves. The simulations have successfully reproduced the
"sea-serpent" behavior of magnetic field lines
\citep{SainzDalda2008,Kitiashvili2010}, and  showed that the Evershed
flow is concentrated in the upper 1-Mm deep layer
\citep{Kitiashvili2009}.

The outflow outside the penumbra (called "moat flow") was first observed in
Doppler velocities of the photospheric plasma \citep{Sheeley1969,Sheeley1972}. It is
associated with an outflow of moving bipolar magnetic features (MMF).
Observations also showed that the sunspot moat flow is
closely related to the Evershed flow, because it is observed only on
the sides of sunspots, which have the penumbrae
\citep{SainzDalda2005,VargasDominguez2007,Sobotka2007,VargasDominguez2008,
MartinezPillet2009}.
On the sides without penumbrae the moat flow is absent, and instead
inflow is observed. Analysis of high-resolution {\it Hinode} data
reinforced the evidence that the moat flow represents an extension
of the Evershed flow beyond the penumbra boundary
\citep{VargasDominguez2010}. However, an unusual case was reported
by \citet{Zuccarello2009} when several moving bipolar magnetic
elements were observed moving away from a sunspot without penumbra. Also,
observations show that the moving magnetic elements are not
passively transported by the moat flow of plasma, and they can move
faster than the plasma \citep{Balthasar2010}. This is consistent
with the sea-serpent behavior of the magnetic-field lines in the
penumbra, which also produces moving bipolar elements, as was
originally suggested by \citet{Harvey1973}. The MMF flow intensifies
around decaying sunspots. Thus, it is likely that it carries away
some of the magnetic flux concentrated in sunspots. The moat flow
has not been reproduced in simulations. Therefore, its nature is
much less clear than the Evershed flow.

Radiative MHD simulations \citep{Rempel2009} have provided a
fairly realistic pictures of the surface structure of sunspots.
However, in subsurface layers the magnetic-field structure is held together by a
boundary condition that fixes the magnetic field concentration at
the bottom of the simulation domain. Once, the boundary condition is
released the sunspot structure is dispersed. Unlike the stable
configurations of \citet{Hurlburt2000}, the subsurface flows in this
model show a diverging rather than a converging pattern. This is
probably the reason for the instability of the magnetic sunspot
structure in Rempel's simulations.

\begin{figure}
\centerline{\includegraphics[width=0.5\textwidth]{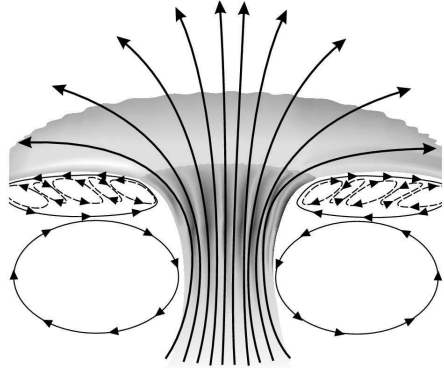}
} \caption{Schematic illustration of flows beneath sunspots, which
shows two patterns of flow circulating in opposite directions: a
shallow Evershed outflows driven by overturning convection and
deeper  converging flows \citep{Zhao2010}.} \label{fig6}
\end{figure}

A complete consistent model of sunspots as self-organized magnetic
structures is not yet developed. However, it becomes clear that the
subsurface dynamics of stable magnetic structures representing pores
and sunspots must include inflows preventing a rapid dispersion
of magnetic field. The Evershed and moat flows are likely to be
quite shallow and driven by the overturning granular convection in
the penumbra regions with almost horizontal strong magnetic field. A
schematic picture of sunspot subsurface flows is given in
Figure~\ref{fig6}. The flow pattern consists of two parts: a shallow
layer of overturning convection in the inclined magnetic field of
penumbra, which provides the mean Evershed outflow, and a deeper
converging (``collar'') flow beneath the Evershed flows.

\section{Local Helioseismology Inferences of Subsurface Flows
and Wave-Speed Structures}

\subsection{Time-Distance Helioseismology}

\begin{figure}
\begin{center}
\includegraphics[width=\linewidth]{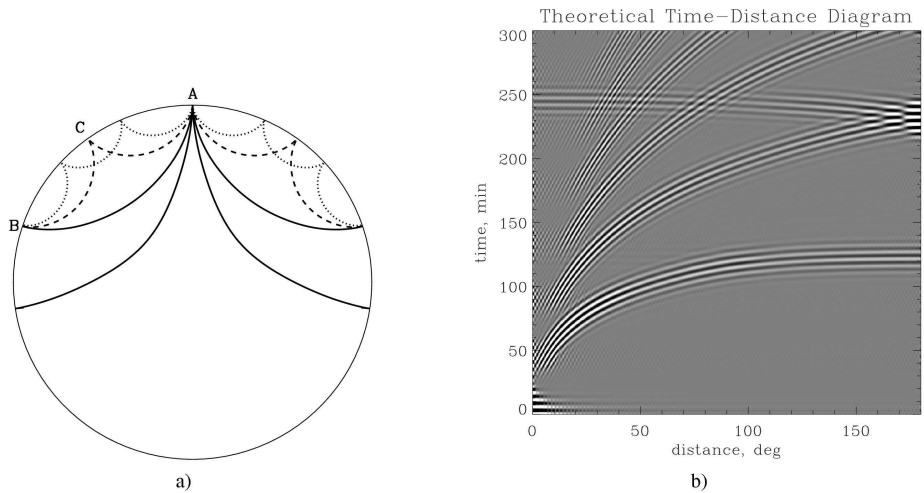}
\caption{a) A sample of ray paths of acoustic waves propagating
through the Sun's interior from surface point A. b) The theoretical two-point
cross-covariance of solar oscillations
as a function of the distance and lag time. The lowest ridge
corresponds to wave packets propagating between two points on the solar
surface directly, {\it e.g.} along ray path AB (solid curve). The second
ridge from below
corresponds to acoustic waves that have an additional
reflection at the surface  (``second bounce''), {\it e.g.} along ray path ACB.
This ridge appears reflected at the distance of 180$^\circ$, because
the propagation distance is measured in the interval from 0$^\circ$ to 180$^\circ$ \citep{Kosovichev2003}.}\label{fig7}
\end{center}
\end{figure}

Historically, the first helioseismology inferences of the structures
and flows beneath sunspots and active regions were made by using
the technique of time--distance helioseismology \citep{Duvall1993}.
This technique is based on measuring travel times of acoustic waves. Solar acoustic waves ({\it p} modes) are excited by turbulent convection near the solar surface
and travel through the interior with the speed of sound. Because the sound speed
increases with depth, the waves are refracted and reappear on the surface at some
distance from the source. The wave propagation paths are illustrated in Figure~\ref{fig7}a.

The basic idea of time-distance helioseismology, or helioseismic tomography, is to measure the
acoustic travel time between different points on the solar surface,
and then to use these measurements for inferring variations of wave-speed perturbations
and flow velocities in the interior by inversion. This idea is similar to seismology of the Earth.
However, unlike in the Earth, the solar waves are generated stochastically
by numerous acoustic sources in a subsurface layer of turbulent
convection. Therefore, the wave travel time and other wave-propagation
properties are determined from a cross-covariance
function of the oscillation signals observed at different points on the solar surface.  A typical
cross-covariance function calculated for the whole disk is shown in
 Figure \ref{fig7}b \citep{Kosovichev2003}. It displays a set of ridges. The lowest ridge
corresponds to wave propagating directly
between two surface points, (the lowest ridge). The second ridge from below
is formed by
waves which experience one additional reflection at the surface on their way
from point A to B,
{\it e.g.} wavepath ACB in Figure \ref{fig7}a, (so-called ``second bounce'' ridge). The upper
ridges correspond to waves with multiple reflections from the surface.
The cross-covariance function represents a `helioseismogram'.

\begin{figure}
\centerline{\includegraphics[width=0.65\textwidth]{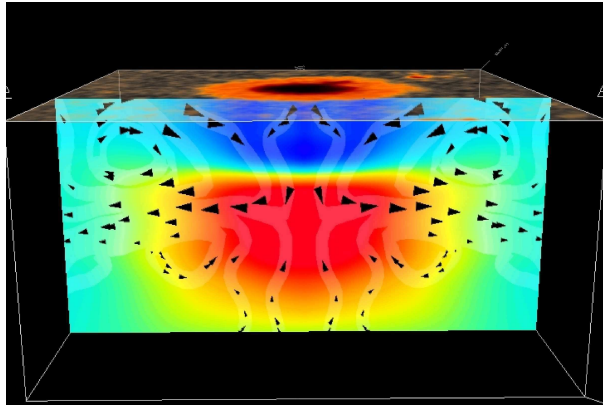}
} \caption{Wave-speed variations and axisymmetric component of mass
flows beneath a sunspot inferred by inversion of acoustic travel
times in the ray-path approximation
(after \citet{Kosovichev2000} and \citet{Zhao2001}).} \label{fig8} \end{figure}

\citet{Duvall1996}, using helioseismic observations from the geographical
South Pole, measured the travel-time difference between the wave
traveling from a sunspot and towards the spot, and they concluded that
this difference could be explained by downward mass flows beneath the
sunspot. \citet{Kosovichev1996} developed a tomographic inversion procedure
for the travel times, based on a ray-theoretical approximation, and
obtained first subsurface maps of the sound-speed variations and flow
velocities. These maps were of rather low spatial resolution ($\approx
16$ Mm), but revealed large-scale subsurface converging flows around
active regions. This technique was then improved by
\citet{Kosovichev1997} and applied to the analysis of helioseismology
data from SOHO/MDI \citep{Kosovichev2000}. The inversion results
showed a two-layer subsurface structure of a sunspot with a negative
wave-speed perturbation in the top 4\,--\,5 Mm layer and a positive
perturbation in a deeper layer (Figure~\ref{fig8}). Their results
also showed the process of formation of the sunspot structure during
the magnetic flux emergence.  The two-layer character of the sunspot
wave-speed structure has been confirmed by \citet{Jensen2001},
\citet{Couvidat2004}, and \citet{Couvidat2006}, who took into account
the finite-wavelength effects by using the Fresnel-zone and Born approximations.
 Similar inversion results
were also obtained by \citet{Zharkov2007}, and most recently by
\citet{Zhao2010} from {\it Hinode}/SOT data.

These results have been a subject of debate for more than a decade
mainly for three
reasons: {\it i}) uncertainties in the travel-time measurements; {\it ii})
theoretical approximations used in the travel-time inversion procedures;
{\it iii}) uncertainties in the
interpretation of the inferred wave-speed perturbations in terms of
the thermodynamic and magnetic structure (because the effects of
temperature and magnetic field are not separated). For instance, the
travel-time measurements can be affected by errors of the
Doppler-shift measurements in regions with strong magnetic field, by
spatial variations of the acoustic power, and by the
use of a phase-speed filter, which was applied for measuring the
acoustic travel times for short distances \citep{Duvall1997}.

The
travel-time measurements for the short distances (0.5\,--\,2 heliographic
degrees) are particularly important for inferring the shallow
subsurface layer of the negative wave-speed perturbations. Qualitatively, the subsurface
structure can be deduced from the travel times without inversion.
Indeed, the travel-time perturbations for the short
distances are positive meaning that the wave speed is reduced. For longer propagating distances,
the perturbations are negative indicating faster wave propagating, and thus an increase of the
wave speed.

Most of the tomographic inversions of travel-time variations have
been done by using the ray-path approximation, and assuming that
changes of the ray path do not significantly affect the
travel-times variations (so-called Fermat's principle)
\citep{Kosovichev1997}. This has been tested by using the first Born
approximation, which takes into account  finite wavelength and
frequency effects \citep{Birch2000,Birch2004a,Couvidat2006c}.

There have been attempts to resolve the issue of interpretation
of the wave-speed inferences through modeling
of the magnetostatic structure of sunspots
\citep{Olshevsky2008,Shelyag2009,Cally2009}. The results are
obviously model-dependent, but seem to indicate that the thermal
effects dominate except, perhaps, in the penumbra region, where the
magnetic field is inclined. It was also noticed by
\citet{Kosovichev2000} that the interpretation of the inferred
wave-speed variations purely in terms of deep magnetic fields would require
a significant increase of the magnetic flux of sunspots with depth,
which may be difficult to explain.

The issue of the phase-speed filtering has been
investigated through modeling, which showed that because of the
suppression of the acoustic power in sunspots the phase-speed
filtering may result in systematic shifts in travel-time
measurements \citep{Rajaguru2006}. However, the systematic errors due
to this effect are relatively small, $\lesssim 10$~s
\citep{Parchevsky2008,Hanasoge2008}, and do not affect the principal conclusions about the sunspot structure.

Using high-resolution
{\it Hinode}/SOT observations \citep{Kosugi2007,Tsuneta2008},
\citet{Zhao2010} have been able to measure the travel times for
short distances without the phase-speed filtering procedure and
confirm the positive travel-time perturbations for the acoustic
waves traveling to short distances in a large-sunspot area in
agreement with the results obtained with the phase-speed filtering.
However, these results showed taht the systematic errors may reach 20\%\,--\,40\%. Thus, the
inferences of the wave-speed structure remain largely on a
qualitative level.  The uncertainties caused by the
phase-filtering procedure are discussed in more detail in Section~\ref{filtering}.

\begin{figure}
\centerline{\includegraphics[width=0.7\textwidth]{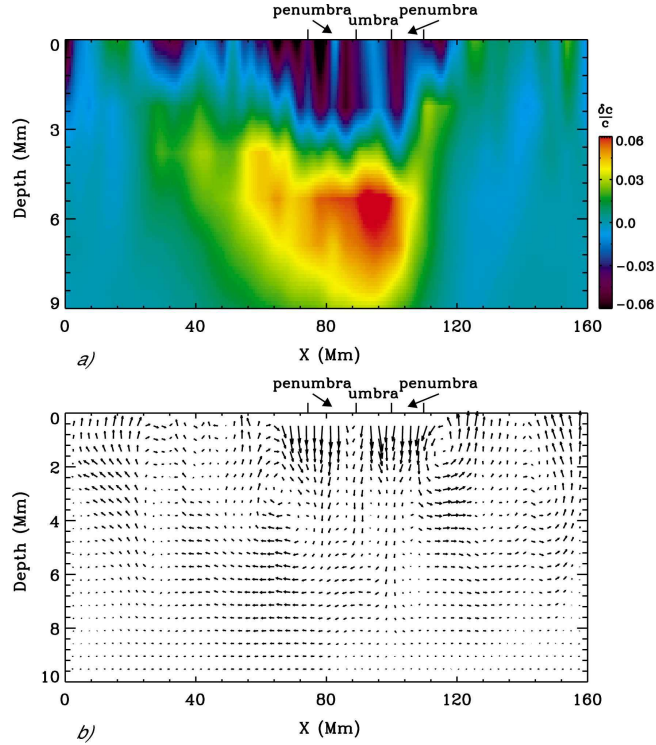}
} \caption{ Vertical cuts through the subsurface wave-speed
structure (a) and the flow field (b) of a large sunspot, obtained
from the time-distance helioseismology data from {\it Hinode}
\citep{Zhao2010}. } \label{fig9} \end{figure}

The tomographic inversion of acoustic travel times measured from the
MDI high-resolution data revealed a converging flow pattern in
the depth range of 1\,--\,5 Mm \citep{Zhao2001}. This  result was
confirmed by analysis of {\it Hinode} helioseismology data
\citep{Zhao2010}. The {\it Hinode} data have provided a more clear and
convincing picture of the converging flow compared to the MDI data
(Figure~\ref{fig9}).

\subsection{Ring-Diagram Analysis}
\begin{figure}
\begin{center}
\includegraphics[width=0.6\linewidth]{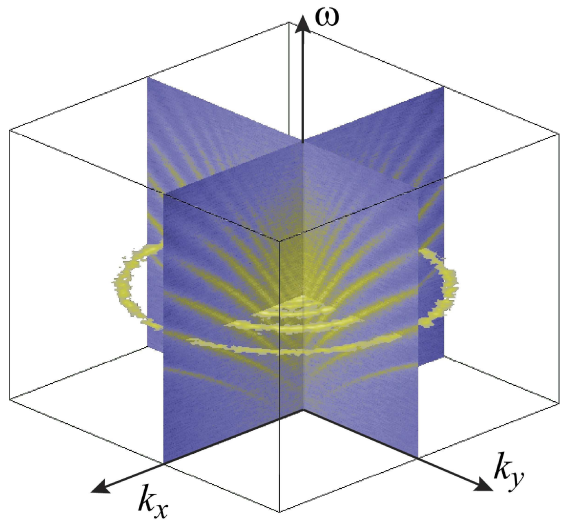}
\caption{Three-dimensional power spectrum of solar oscillations,
$P(k_x,k_y,\omega)$. The vertical panels with blue background
show the mode ridge structure similar to
the global oscillation spectrum.
The horizontal cut with transparent background shows the ring structure
of the power spectrum at a given frequency (courtesy of Amara Graps).}\label{fig10}
\end{center}
\end{figure}

\citet{Gough1983}  proposed to
measure oscillation frequencies of solar modes as a function of the
wavevector (the dispersion relation) in local areas,
and use these measurements for diagnostics of the local flows and thermodynamic properties.
They noticed that subsurface variations of temperature cause change
in the  frequencies, and that subsurface flows result in distortion of
the dispersion relation because of advection.
This idea was implemented by \citet{Hill1988} in the form
of a  ring-diagram analysis. The name of this technique comes from
the ring appearance of the 3D dispersion relation,
$\omega=\omega(k_x,k_y)$, in the $(k_x, k_y)$ plane, where $k_x$ and
$k_y$ are $x$- and $y$-components of the wave vector (Figure~\ref{fig10}). The ridges in the vertical cuts correspond to the normal oscillation modes of different radial orders $n$.

The ring-diagram method has provided important results about
the structure and evolution of large-scale and meridional flows and dynamics of active regions
\citep{Haber2000,Haber2002,Haber2004,Howe2008,Komm2008}. In
particular, large-scale patterns of subsurface flows converging around
magnetic active regions were discovered \cite{Haber2004}. These
flows cause variations of the mean meridional circulation with the
solar cycle \citep{Haber2002}, which may affect transport of magnetic
flux of decaying active regions from low latitudes to the polar
regions, and thus change the duration and magnitude of the solar
cycles.

However, the ring-diagram technique in the present formulation has
limitations in terms of the spatial and temporal resolution and the
depth coverage. The local oscillation power spectra are typically calculated for regions with horizontal size covering 15 heliographic degrees ($\approx 180$ Mm). This is significantly larger than the typical size of supergranulation and active regions ($\approx 30$ Mm). There have been attempts to improve the resolution by doing the measurements in overlapping regions (so-called "dense-packed diagrams"). However, since such measurements are not independent, and the actual resolution is unclear. The measurements of the power spectra calculated for smaller regions (2\,--\,4 degrees in size) can increase the spatial resolution but decrease the depth coverage \citep{Hindman2006}.

\subsection{Comparison of Local Helioseismology Results}
\begin{figure}
\centerline{\includegraphics[width=0.7\textwidth]{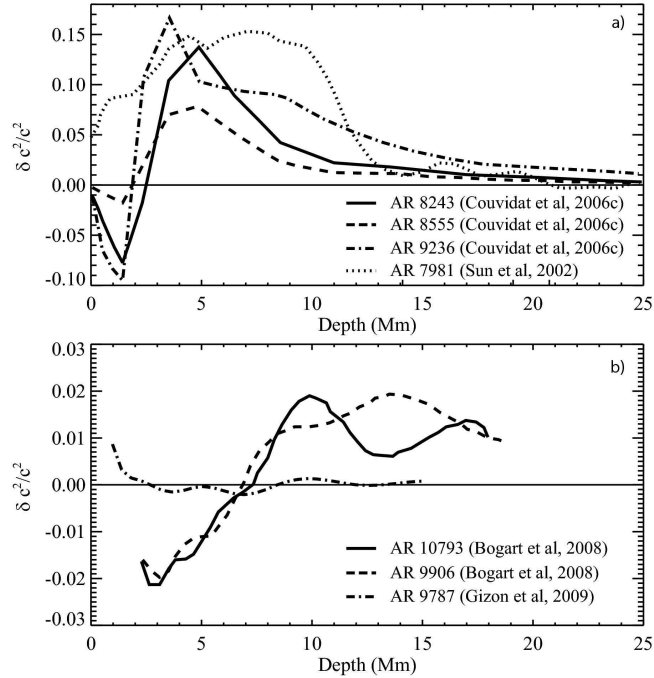}
} \caption{Sound-speed beneath various sunspots obtained by
different local-helioseismology methods: a) time-distance
helioseismology of \citet{Couvidat2006a} (solid, dashed and
dot-dashed curves) and acoustic imaging  of \citet{Sun2002} (dots);
b) the ring-diagram analysis of \citet{Bogart2008} (solid and dashed
curves) and of \citet{Gizon2009} (dot-dashed curve).} \label{fig11}
\end{figure}

The subsurface structure of sunspots have been inferred by three
different local helioseismology techniques: time-distance helioseismology \citep{Kosovichev1996,Kosovichev2000,Zhao2001,Couvidat2006a}
acoustic imaging
\citep{Sun2002}, and ring-diagram analysis
\citep{Basu2004,Bogart2008,Gizon2009}.  We compare
these inferences for various active regions in Figure~\ref{fig11}. In such
comparison it is important to notice that the ring-diagram analysis
has a substantially lower spatial and temporal resolution than the
time-distance helioseismology and acoustic-imaging techniques. The
spatial resolution of the time--distance inversions is typically 3\,--\,6
Mm, and the typical temporal resolution is eight hours, while for the
ring-diagram analysis the typical spatial resolution is about 180
Mm, and the temporal averaging is done at least for 24 hours. This
means that the ring-diagram results are averaged over a large area
covering not only the sunspots, but the whole active region including
plages and also surroundings. Thus, the ring-diagram results cannot be directly compared
with the inferences for sunspots.
The subsurface structure of plages is obviously different from sunspots. Also, the ring-diagram
results are not simple averages of the subsurface perturbations
caused by sunspots and plages, but these averages are
weighted according to the acoustic-power distribution.
Since the acoustic power is
suppressed in sunspots and enhanced (particularly at high
frequencies) in plages, it is likely that the contribution of
sunspots in these inferences is smaller than this can be expected
from a simple averaging. Nevertheless, it is interesting to make a
qualitative comparison of the inversion results obtained by the
different techniques.

We illustrate this comparison in Figure~\ref{fig11}. The top panel
shows the results of the time-distance helioseismology and acoustic
imaging analyses for four sunspot regions. The sound-speed
variations represent averages over the central part of sunspots
relative to the surrounding quiet-Sun values. The common feature of
these results is an enhancement in the deep interior 2--10 Mm, and a
decrease in the subsurface layers. The time-distance results show a
shallow negative variation, while the acoustic imaging inversion
does not show this. The results of the ring-diagram inversions,
obtained by \citet{Bogart2008}, and also by \citet{Basu2004}, show
the structure qualitatively similar to the time-distance results,
with a positive sound-speed variation in the deep interior and a
negative variation in the subsurface layer, but this layer is deeper
than in the time-distance profiles. Recently, these results were
confirmed by a statistical study of \citet{Baldner2009}, who have
found the sound speed depressed in the near surface layers, but enhanced below that,
and that this variations correlate with a magnetic activity index.

Surprisingly, the results published by \citet{Gizon2009} are
drastically different from the other results obtained by the same
technique. This difference could be due to an unusual structure of
AR 9787, but this is unlikely because the time--distance inversions
for this region according to \citet{Gizon2009} are similar to those
shown in the top panel of Figure~\ref{fig11}. The reasons for this
difference are unclear, and currently being investigated in
detail. Without such investigation it is premature to conclude that
this inconsistency shows a failure of the local helioseismology
inversions, as this was suggested by
\citet{Gizon2009}. Based on these results, \citet{Moradi2010} suggested 
that the sunspot in AR 9787 is most probably associated with a shallow, 
positive wave-speed perturbation (unlike the traditional two-layer model).

However, further investigation of the same active region (NOAA 9787) by 
\citet{Kosovichev2010} showed  that the inversion results obtained by two different methods of local helioseismology, the ring-diagram analysis and time-distance helioseismology, are consistent with most of the previous results for other active regions, revealing the characteristic two-layer structure with a negative variation of the sound speed in a shallow subsurface layer and a positive variation in the deeper interior. However, there are significant quantitative differences between the inversion results obtained by the different techniques and different inversion methods. In particular, the seismic structure of the active region inferred by the ring-diagram method appears more spread with depth than the structure obtained from the time-distance technique.

It was also pointed out that the quantitative comparison of the inversion results is not straightforward because of the substantially different spatial resolutions of the helioseismology methods. The quantitative comparison must take into account differences in the sensitivity and resolution. In particular, because of the acoustic power suppression the contribution of the sunspot seismic structure to the ring-diagram signal can be substantially reduced. \citet{Kosovichev2010} showed that taking into account this effect  reduces the difference in the depth of the sound-speed transition region. 
Their results obtained by the two local helioseismology methods indicate that the seismic structure of sunspots is probably rather deep, and extends to at least 20 Mm below the surface. If confirmed by further studies this conclusion has important implications for development of theoretical models of sunspots.

A systematic comparison of the subsurface flow patterns beneath
sunspots and active regions has not been done. However, both, the
time--distance and ring-diagram helioseismic techniques have shown
remarkably similar results for large-scale subsurface flows
\citep{Hindman2003,Hindman2004}, with common inflow sites around
active regions as well as agreement in the general flow direction.
At a depth of $\approx 1.5$ Mm the correlation coefficient between the maps
is about 0.80. As the depth increases the correlation
becomes weaker. The reduction in the correlation coefficient with
depth may be due to the increasing difference between the vertical
resolution kernels of these techniques. Also, the recent ring-diagram
inferences based on measurements of {\it f}-mode frequency shifts with a
higher resolution ($\approx 25$ Mm) revealed the near-surface outflows in
the moat-flow region of sunspots \citep{Hindman2009}. This is
consistent with the time--distance results also from {\it f}-mode measurements
\citep{Hindman2004}. Based on the ring-diagram results,
\citet{Hindman2009} suggested that sunspots are surrounded by a
shallow, less than 2 Mm deep, moat flows and by deeper converging
large-scale flows. Because of the low resolution, the structure of
the converging flows on the scale of sunspots cannot be assessed,
but these inferences are not inconsistent with the time-distance
results inferred from the {\it f}- and {\it p}-mode travel time measurements.
Similar results in the moat flow region, obtained by a new
ridge-filtering approach to time-distance helioseismology, have been
reported by \citet{Gizon2009}. However, these results indicate that
the moat outflow may be extended into the deeper layers with no sign
of reversed flows. The reality of such one-directional flow over
a large range of depth is difficult to
assess also because these measurements were based on a
cross-covariance linearization method \citep{Gizon2004}, which may
give significant systematic errors in sunspot regions,
as pointed out by \citep{Couvidat2010} (Section~\ref{definitions}).

Thus, in general, the theoretical picture for mature stable
sunspots, illustrated in Figure~\ref{fig8}, is  consistent with the
results of local helioseismology obtained by the time--distance
technique \citep{Zhao2001,Zhao2010} and by the ring-diagram analysis
\citep{Haber2004,Hindman2009}. The measurements of the frequency
shifts and travel times of the surface gravity waves ({\it f} modes) and
acoustic waves ({\it p} modes) show opposite signs, consistent with an
outflow in a shallow subsurface region and an inflow in the deeper
interior. The surface gravity waves travel in a relatively thin
subsurface layer and thus are sensitive to the properties of this
layer, while the acoustic waves travel much deeper. However,
these parts of the flow field beneath sunspots have been
inferred by helioseismic inversions separately by inversion of the
{\it f}- and {\it p}-mode travel times. An unified flow circulation
pattern in sunspots has not been obtained. Developing an inversion
procedure for combined {\it f}- and {\it p}-mode travel-time data is a very important task.

The current investigations also include numerical
simulations of helioseismic data with flows and sunspot models for
verification and testing the inferences, investigations of various
uncertainties and development of new methods of local
helioseismology of sunspot regions.

\subsection{Changes of Subsurface Structures and Flows During
Growth and Decay of Sunspots}

The structure and dynamics of sunspots change substantially during their formation and evolution. A very important helioseismology task is to detect signature of the magnetic flux of active regions before it becomes visible on the surface and forms sunspots. However, this task turned out be difficult because the emerging magnetic flux travels very rapidly in the upper convection zone, with a speed exceeding 1~km/s \citep{Kosovichev2000}. Thus, it takes less than 8 hours for the flux to emerge form the depth of 30~Mm, and the typical observing time required for measuring travel times is also eight hours. The time series for measuring frequency shifts by the ring-diagram technique are 24 hours or longer. An attempt to detect emerging flux using short two-hour time series of Doppler-shift observations of solar oscillations from SOHO/MDI was made by \citet{Kosovichev2000}. These data revealed a signature of emerging flux in the sound-speed images beneath the surface. However, no indication of emergence of a strong large-scale $\Omega$-loop predicted by theories has been obtained. In addition, surface Doppler-shift observations of a large emerging active region revealed strong localized upflows and downflows at the initial phase of emergence but found no evidence for large-scale flows indicating future appearance of a large-scale magnetic structure \citep{Kosovichev2009b}.  It seems that the active regions are formed over extended period of time as a result of multiple magnetic flux emergence events. The results of time--distance helioseismology showed predominantly  diverging flow patterns during the magnetic-flux emergence and decay, and mostly converging flows around stable sunspots at  1\,--\,4 Mm depth \citep{Kosovichev2006,Kosovichev2009b}. The derived vertical-flow pattern is complicated during flux emergence with intermittent up- and downflows. However, on average, the upflows are dominant at the beginning of the emergence phase, but then replaced by downflows when sunspots are developed. The flow divergence shows a correlation with flux-emergence events during the evolution of a large active region (AR 10488), but it is unclear if the flux emergence rate precedes the variation of the flow divergence or follows it.

The ring-diagram inversion results \citep{Komm2007,Komm2009a,Komm2009} are generally consistent with the time--distance helioseismology inferences. The vertical velocity is not measured by this method, but estimated from the horizontal velocity pattern using a stationary continuity equation and assuming that the density stratification is horizontally uniform and corresponds to a quiet-Sun model. The accuracy of these assumptions for the case of non-stationary and non-uniform structure and dynamics of active regions has not been tested. However, these estimates show that the subsurface upflows are stronger for stronger emerging flux, and that the flows change to downflows after the active regions are developed.

\section{Uncertainties in Local Helioseismology Inferences}

\subsection{Uncertainties of Doppler-shift Measurements}

Most local-helioseismology inferences have been carried out by using solar
oscillation velocity data obtained by measuring the Doppler shift of
spectral lines formed in the solar photosphere. In particular, the
SOHO/MDI and GONG measurements are obtained by observing the Ni~{\sc i}
(6768 \AA) line.
In regions of strong
magnetic field the shape of the line is affected by the Zeeman
splitting and other polarization effects.  In the
observations the Doppler shift is calculated by averaging the left-
and right circular polarized components. In the magnetic field,
these components become broader because of the splitting. This
results in an underestimation of the Doppler shift in the sunspot
umbra when a sunspot is located in the central part of the solar
disk, and in the penumbra when the sunspot is near the solar limb.
The uncertainties in acoustic travel times caused by the line
broadening and other radiative transfer effects in the sunspot
atmosphere were investigated by
\citet{Wachter2006a}, \citet{Wachter2006}, and \citet{Rajaguru2007}, who found that around 3 mHz used for helioseismology, the systematic
errors do not exceed five seconds, which is by an order of magnitude lower
than the typical observed travel-time anomalies. Thus, these effects
do not play significant role, but should be taken into account for
improving the precision of the measurements. \citet{Wachter2006a}
suggested a correction procedure for the Doppler-shift measurements in strong-field regions.

\subsection{Inclined-field ("Shower-glass") Effect}

It has been noticed by \citet {Schunker2005} that when a sunspot is
located near the limb the phase shifts of acoustic waves
(corresponding to travel times) vary in a sunspot penumbra
 relative to the direction to the disk center.
They attributed this to a phase change of acoustic waves traveling
through the inclined magnetic field of the penumbra, calling these
phase perturbations ``the acoustic showerglass'', and suggesting that
this effect might substantially affect the inversion results.

\begin{figure}
\centerline{\includegraphics[width=0.7\textwidth]{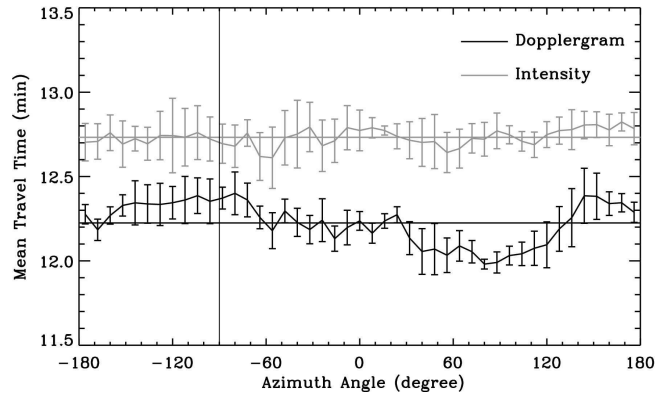}
} \caption{Mean acoustic travel-time variations inside a sunspot
penumbra versus the azimuthal angle, measured from MDI Dopplergrams
and intensitygrams for a sunspot inside AR8243 on 18\,-–\,19 June 1998.
The dark curve represents results computed from the Dopplergrams,
and the light gray curve represents results from the continuum
intensitygrams. The error bars are standard deviations. The dark and
gray horizontal lines indicate the average mean travel time inside
the penumbra from the Dopplergrams and intensitygrams, respectively.
The vertical line indicates the azimuthal angle of the solar disk
center relative to the center of the sunspot \citep{Zhao2006}. }
\label{fig12} \end{figure}

\citet{Zhao2006} investigated this effect in detail by using the
standard time--distance helioseismology procedure. They qualitatively reproduced
this effect for the travel times measured from the Doppler-shift
data; but found that this effect is completely absent in the travel
times obtained from the MDI intensity-oscillation data
(Figure~\ref{fig12}). This means that the inclined-field effect is
probably caused by the radiative transfer effects in a magnetic field,
affecting the Doppler-shift measurements, rather than by changes in
wave-propagation properties due to the surface magnetism. There is also a
possibility that the Doppler-shift signal might be affected by
changes in the relationship between the vertical and horizontal
components of the oscillation in the inclined field regions. This
needs to be investigated.

\begin{figure}
\centerline{\includegraphics[width=0.95\textwidth]{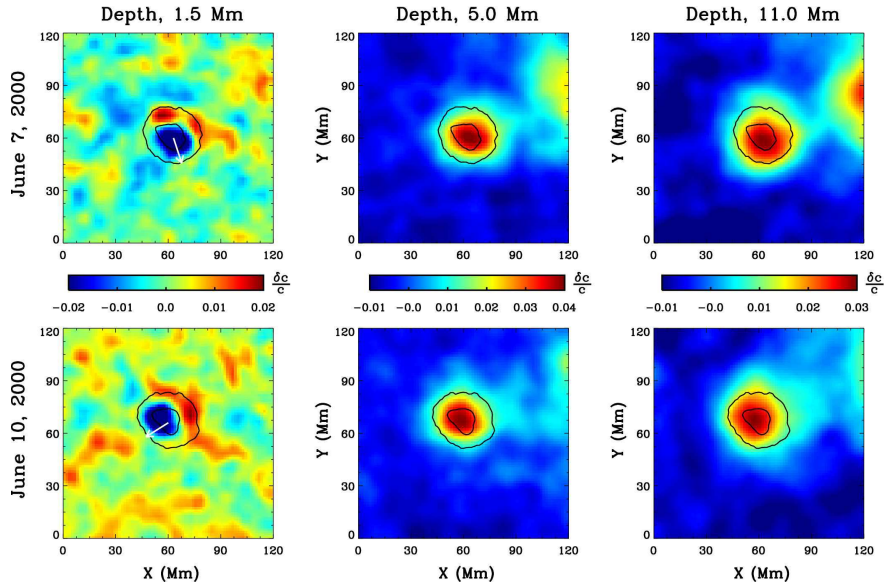}
} \caption{Sound-speed variations inferred from time--distance
inversions for AR9026, shown at selected depths: 1.5~Mm (left),
5.0~Mm (middle), and 11.0~Mm (right), for two dates: 7 June 2000
(top) and 10 June (bottom), when the distance from the disk center
was 70 and 150 degrees respectively. Contours indicate the
boundaries of the sunspot umbra and penumbra determined from MDI
continuum-intensity observations. White arrows at a depth of 1.5~Mm
on both dates point to the solar disk center. For different dates,
the image display color index is the same for the same depth, with
the color bars shown in the middle row \citep{Zhao2006}.}
\label{fig13} \end{figure}

Since most helioseismic observations are based on
Doppler-shift data, \citet{Zhao2006} investigated the systematic
errors in the sound-speed inversion results caused by the
inclined-field effect by comparing the inversion results for
different positions of a sunspot on the solar disk. The results
(Figure~\ref{fig13}) showed that that the perturbations of the travel
times cause a systematic shift of the sound-speed variations in the
near-surface layers (1.5 Mm deep), but do not affect the inversion
results in the deeper layers.

For the local helioseismology of sunspots, it is important that such
an effect is absent in the intensity oscillation data, and that the
analysis of the intensity data from SOHO/MDI and {\it Hinode}
(Figure~\ref{fig9}) has confirmed the basic inferences obtained from
the Doppler-shift data. However, for improving the precision of the
local helioseismic diagnostics from Doppler-shift data it is important to
develop a procedure for correcting the travel-time perturbations in the
inclined-field regions of penumbrae. Also, for better understanding the
physics of this effect has to be investigated by forward
MHD modeling ({\it e.g.} \citet{Parchevsky2009}).

\subsection{Effects of Phase-Speed Filter and Acoustic Power Suppression}
\label{filtering}

\begin{figure}
\centerline{\includegraphics[width=\textwidth]{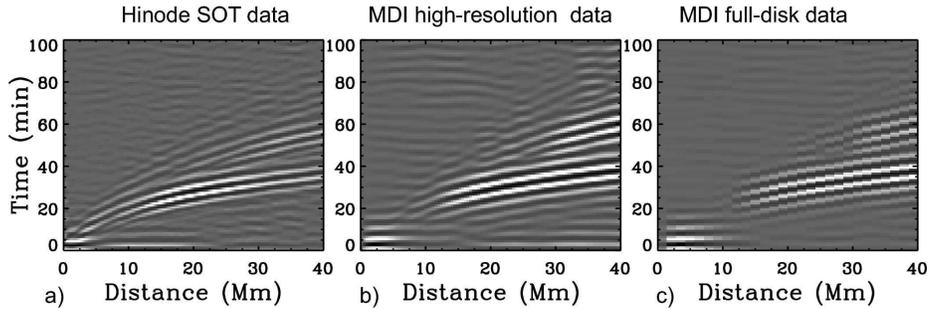}
} \caption{Time–-distance diagrams obtained from a) {\it Hinode} Ca~\sc{ii}~H intensity images; b) MDI high-resolution Dopplergrams; c) MDI full-disk Dopplergrams \citep{Kosovichev2009c,Zhao2010}. } \label{fig14} \end{figure}

The systematic errors caused by a phase-speed filter applied
to the MDI data in order to
improve the signal-to-noise ratio at short travel distances are a concern \citep{Birch2009}. The
measurements of the acoustic travel times for short travel distances
of 0.3\,--\,0.8 heliographic degrees (4\,--\,10 Mm) are necessary for
inferring the structure of the shallow subsurface layers of
sunspots. However, the time--distance diagrams obtained from the MDI
data are corrupted by a set of horizontal ridges\,--\,artifacts, probably, due to instrumental effects
and leakage of low-degree oscillations (Figure~\ref{fig14}b--c). In the high-resolution intensity data from {\it Hinode} (Figure~\ref{fig14}a), the horizontal artifact ridges are much weaker than in the MDI data.
Phase-speed filtering
was introduced to separate the acoustic-wave signal from the
artifact in the MDI data by \citet{Duvall1997}. The phase-speed filter is set to select
the signal corresponding to the select the waves traveling to a
particular range of distances using a theoretical relationship between the
wave's horizontal phase speed and the travel distance.

Phase-speed
filtering substantially improves the signal-to-noise ratio for the
short-distance measurements. However, because of the strong local reduction of the oscillation power in
sunspots, the phase-speed filtering causes a systematic shift in the
travel times measured following the original time--distance
helioseismology procedure of \citet{Kosovichev1997}. Their
formulation of the travel-time Gabor-wavelet fitting formula did not
include the phase-speed filtering.  Phase-speed filtering was
included in the theory only recently by \citet{Nigam2010}, who
derived a new fitting formula. However, because of complexity, this
formula has not been used. The travel-time shift in the region of
acoustic-power reductions was found empirically by
\citet{Rajaguru2006}. This effect was modeled by
\citet{Hanasoge2008} and \citet{Parchevsky2008}, who found that the travel-time
shifts are only a few seconds (reaching $\approx 15$ seconds in an extreme
case of the \citet{Hanasoge2008} simulations), and do not significantly change
the inversion results. Generally, this effect causes
underestimation of the sound-speed variations in the shallow
subsurface layers. The shift can be reduced by normalizing the power
variations.

\begin{figure}
\centerline{\includegraphics[width=0.95\textwidth]{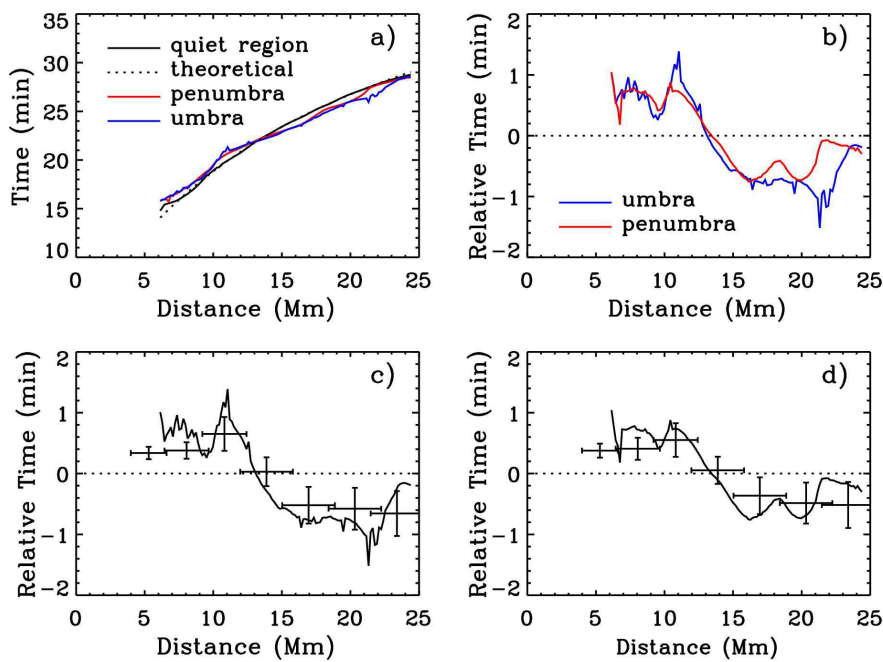}
} \caption{(a) Comparison of the acoustic travel times obtained for a quiet
region (solid black curve), sunspot penumbra (red), and sunspot umbra
(blue) without phase-speed filtering. The dotted line is an estimate
from  ray-path theory. (b) Travel-time differences relative to
the quiet region for the sunspot penumbra and umbra without the phase-speed filtering.
(c) Comparison of
the travel-time differences without (curve) and with (points with errorbars)
phase-speed filtering for the
sunspot umbra. (d) Same as panel (c), but for the penumbra.
Horizontal bars indicate the range of distances;
and the vertical bars indicate the standard errors.
\citep{Zhao2010}.} \label{fig15} \end{figure}

Using the {\it Hinode} observations made with a  {\it Solar Optical
Telescope}, \citet{Zhao2010} were able to obtain the travel-time
measurements for  short distances without  phase-speed
filtering and confirm the sound-speed results,
obtained from the MDI data with the phase-speed filtering
(Figure~\ref{fig15}).

However, the variations of acoustic power in sunspot regions may
have significant effects on inferences of subsurface flows, because
the suppression of acoustic sources in sunspots causes anisotropy in
wave propagation properties deduced from the cross-covariance
function. This must be carefully investigated using numerical
simulations.

\subsection{Cross-talk Effects}

It is important to note that the measurements of the vertical flows by the time--distance technique may have systematic errors due to a cross-talk effect. Because of the specific geometry of the acoustic wave paths, a regular pattern of a horizontal diverging flow may give an artificial downflow contribution to the vertical-velocity estimates \citep{Kosovichev1997,Zhao2003a}. For instance, a horizontal outflow from point A in Figure~\ref{fig7}a will accelerate the waves traveling from this point in a similar way as a downflow at this point. In some cases, when the horizontal flow divergence is strong but the vertical flow is weak the inversion results for the vertical flow may give the incorrect sign. For instance, in supergranulation, where the vertical flow is very weak, the cross-talk gives an artificial downflow signal in the middle of supergranules. However, beneath the sunspots the horizontal flow is converging and the vertical flow is directed downward (Figure~\ref{fig8}). Thus, the cross-talk effect results in an underestimation of the downflow speed, but it cannot cause a reversal in the direction of the measured vertical flows.

The role of the cross-talk effect in the travel-time inversions carried out by the LSQR method \citep{Kosovichev1997} has been studied by \citet{Zhao2003a}.  It has been shown that the cross-talk may be significant when the inversion results are obtained with a small number of iterations (five,\--\,ten) in the LSQR algorithm, which is typically required for inversion of noisy data. However, the inversion with a large number of iteration ($\approx 100$) substantially reduces the cross-talk and gives the correct answer even for weak vertical flows. However, this requires reducing the noise level in the travel-time measurements ({\it e.g.} by increasing the measurement time). For improving the diagnostics of sunspots, it is important to develop an inversion  procedure specifically minimizing the cross-talk effect \citep{Jackiewicz2009}.

\subsection{Travel-time Definitions}
\label{definitions}

Helioseismic travel times are measured by using the
Gabor-wavelet fitting formula derived by \citet{Kosovichev1997}. It
has certain limitations because it was derived assuming the uniform
distribution of acoustic sources and did not include phase-speed
filtering. The phase and group travel times are measured by fitting
this formula to the calculated cross-covariance function using a
least-squares minimization procedure. \citet{Gizon2002,Gizon2004}
adapted two other procedures originally developed in geophysics. The
first is based on minimization in terms of least-squares of the
difference between the observed cross-covariance function and a
reference cross-covariance function, which can be theoretical or
calculated for a quiet-Sun region. The second procedure was based on
a linearization of this difference.

\begin{figure}
\centerline{\includegraphics[width=0.98\textwidth]{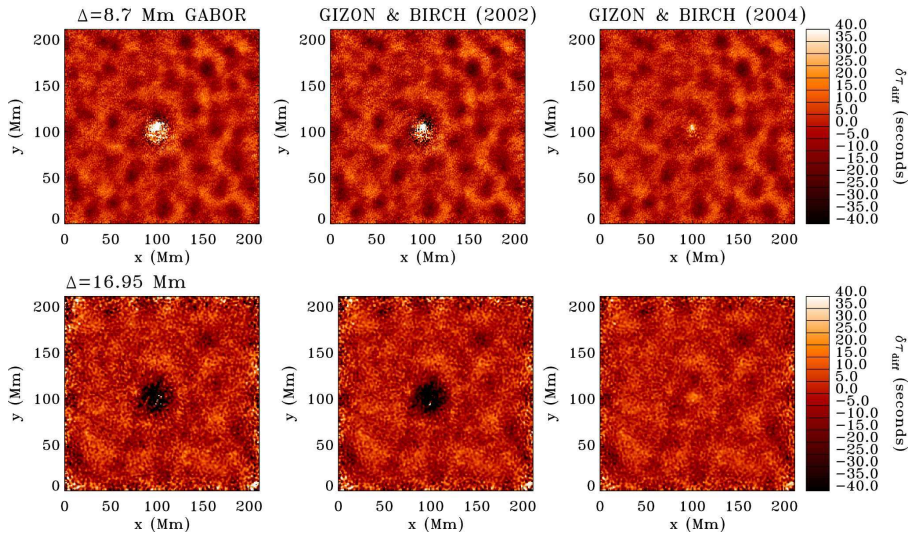}
} \caption{Difference travel-time perturbations for the active
region NOAA 8243 obtained by using methods of fitting the Gabor
wavelet \citep{Kosovichev1997} (left column), minimizing the
difference between the cross-covariance functions calculated for
sunspot and quiet-Sun regions \citep{Gizon2002} (central column),
and by linearizing this difference \citep{Gizon2004} (right panel),
and for two source-receiver distances \citep{Couvidat2010}.}
\label{fig16} \end{figure}

Recently, \citet{Couvidat2010} conducted extensive comparison of
these procedures, and found that the travel times measured by
the approaches of \citet{Kosovichev1997} and \citet{Gizon2002} provide very similar
results, but the linearization approach \citep{Gizon2004} gives
significantly different travel times (Figure~\ref{fig16}). The reason for this discrepancy is
probably in the strong variations of the acoustic power in sunspots,
which are not accounted for in the linearization algorithm.
In  quiet-Sun regions, all three methods give consistent results.
It was concluded that the use of the travel-time definition of
\citet{Gizon2004} in sunspot regions is problematic. This causes
concerns about the inferences of subsurface flows reported by
\citet{Gizon2009} based on this definition.

\section{Interaction of Helioseismic Waves with Sunspots}

Local helioseismic inferences are based on simple models based
on basic principles of wave propagation and physical intuition. A
very important role in verification and testing of these results is
played by numerical 3D MHD simulations, which became possible in recent
years. Substantial progress is also being made in modeling
sunspots. In addition, for understanding sunspot seismology it
is important to study the physics of the wave interaction with a
sunspot using forward modeling. A fundamental characteristic of
wave physics and seismology is the wave Green's function, which
models the waves excited by elementary point sources. On the Sun
such localized sources are provided by solar flares. However, in
most cases the flare sources are anisotropic and moving, thus
producing waves of complex shape and characteristics
\citep{Kosovichev2006a}.

\begin{figure}
\centerline{\includegraphics[width=0.85\textwidth]{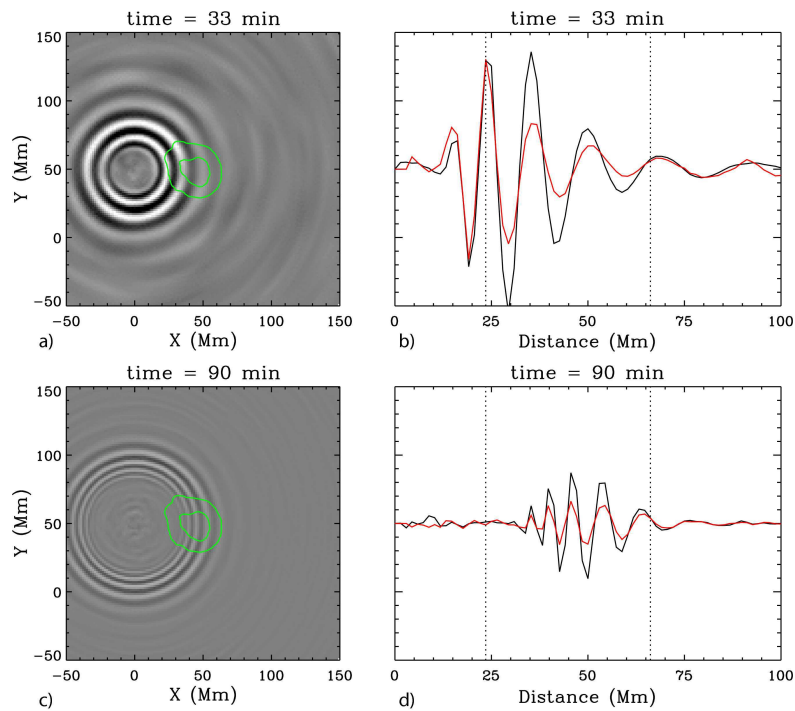}
} \caption{Averaged cross-correlation functions showing the
interaction of acoustic ({\it p}-mode) waves (panels a) and surface-gravity waves ({\it f}-mode) waves (panels c) from an effective point
source with a sunspot, at the moments when the wavefronts cross the
sunspot. Panels b) and d) compare the waveforms profiles along the
$x$-axis (red curves) with the corresponding waveforms calculated
for a quiet-Sun region. The vertical lines indicate the positions of
the sunspot boundaries \citep{Zhao2010a}.} \label{fig17}
\end{figure}

In the case of stochastically excited waves, an effective Green's
function is represented by the two-point cross-covariance function. With
sufficient averaging in time and space this function can be visualized
and compared with the results of numerical simulations of MHD waves
from point sources, calculated for various sunspot models.
Figure~\ref{fig17} illustrates the averaged cross-correlation
functions for acoustic ({\it p}-mode) and surface gravity ({\it f}-mode) waves
traveling through a sunspot \citep{Zhao2010a}. These functions show
how the amplitude and phase of these waves change when they travel though
this spot. These changes, particularly, of the wave phase  depend
also on the position of the source relative to sunspot. The results
show that the {\it f}-mode waves are affected by the sunspot
significantly more than the {\it p}-mode waves. The
{\it p}-mode waves recover their amplitude after passing through the
sunspot, because they travel through the deep interior interior
 where perturbations are small, while the amplitude of {\it f}-mode waves remains reduced.

\begin{figure}
\centerline{\includegraphics[width=0.7\textwidth]{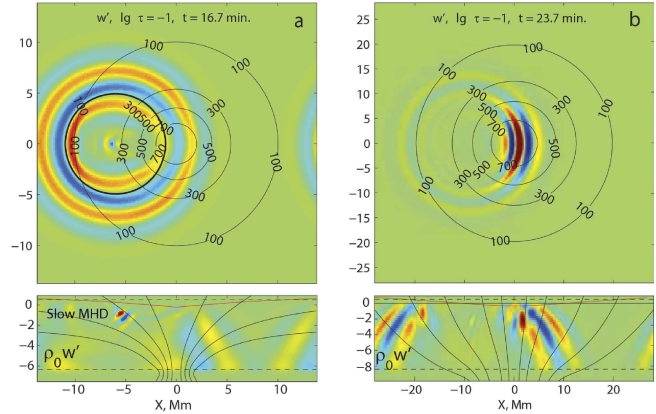}
} \caption{Snapshots of the vertical component of velocity of the
MHD waves, excited by point sources and traveling through two
sunspot models of \citet{Khomenko2008}: a) "deep" model and b)
"shallow" model. Each panel consists of two pictures: nearly
horizontal slice of the domain at the level $\log\tau = -1$ (top)
and the vertical cuts of the domain (bottom).  Solid black curves with numbers indicating field strength
represent the magnetic
field lines. The black
horizontal line and red curve in the vertical cuts (bottom panels) represent the
position of the quiet photosphere and the level of plasma parameter
$\beta = 1$ respectively. \citep{Parchevsky2010}.} \label{fig18} \end{figure}

These results can be qualitatively compared with the simulations of
MHD waves for two sunspots models of \citet{Khomenko2008}, obtained
by \citet{Parchevsky2010}. The simulation results  (Figure~\ref{fig18}) show
that in a "deep" sunspot model the wave amplitude is reduced similar
to the observations. In a "shallow" model, the amplitude is
substantially increased. This is inconsistent with observations,
and rules out the "shallow" sunspot model.  Thus, such a forward modeling approach allows
us to discriminate among  sunspot models. Similar approach for an
effective source extended in one direction has been recently presented by
\citet{Cameron2010}.

There is no doubt that the forward-modeling approach will be further
developed and used for understanding the physics of the interaction
of helioseismic waves with sunspots. The forward modeling can help
in determining the relative role of magnetic and thermal effects
in the wave-speed variations deduced by helioseismic inversions
\citep{Olshevsky2008}. Also, there is a potential for developing the
wave-form tomography of sunspots as pointed out by \citet{Zhao2010a}.

\section{Conclusion}

Local helioseismology has provided the first important insight into the
subsurface structure and dynamics of sunspots that are the key elements of
the Sun's magnetic activity. A list of the initial inferences is presented
in Table~\ref{table1}. However, because of complexity of the
filamentary, turbulent and dynamic nature of sunspots there are
significant uncertainties in the helioseismic inferences, which
require further investigation (Table~\ref{table2}).

\begin{table}[b]
\caption{Initial local helioseismology inferences by Time--Distance Helioseismology (TDH), Ring-Diagram Analysis (RDA), and Acoustic Imaging (AI)} \label{table1}
\begin{tabular}{l l l}     
  \hline                   
 Results & Method & References \\
  \hline
Downflows under sunspots & TDH & \citet{Duvall1993} \\
Sound-speed increase beneath active regions & TDH & \citet{Kosovichev1996}\\
 & AI & \citet{Sun2002}\\
 & RDA & \citet{Basu2004}\\
 Sound-speed decrease in a shallow  & TDH & \citet{Kosovichev2000}  \\
 subsurface layer & RDA & \citet{Basu2004} \\
 Large-scale converging flows around AR & TDH & \citet{Kosovichev1996} \\
 & RDA & \citet{Haber2004,Komm2007} \\
 Subsurface moat outflow & TDH & \citet{Gizon2001}\\
 Emerging flux and formation of AR & TDH &  \citet{Kosovichev2000};\\
 &  &  \citet{Kosovichev2009b}\\
 & RDA &  \citet{Komm2008}\\
  Vortexes under rotating sunspot & TDH & \citet{Zhao2003}\\
 Cluster structure of a large sunspot & TDH & \citet{Zhao2010}\\
  \hline
\end{tabular}
\end{table}
\begin{table}
\caption{Uncertainties and tests of local helioseismology diagnostics}\label{table2}

\begin{tabular}{p{5.50cm} p{5.50cm}}
\hline
 \multicolumn{1}{p{5.50cm}}{\raggedright Uncertainties and concerns} &  \multicolumn{1}{p{5.50cm}}{\raggedright Potential effects and solutions} \\
\hline
 \multicolumn{1}{p{5.50cm}}{\raggedright Calibration of Doppler shift in strong field regions. \citep{Wachter2006}} &  \multicolumn{1}{p{5.50cm}}{\raggedright Travel-time shifts $\approx five$ seconds at 3 mHz. Correction procedure developed. \citep{Wachter2006a}}
 \vspace*{1mm}\\

 \multicolumn{1}{p{5.50cm}}{\raggedright Phase-speed filtering and acoustic power suppression. \citep{Rajaguru2006}} &
 \multicolumn{1}{p{5.50cm}}{\raggedright Travel-time shifts up to ten~seconds. Correction procedure developed \citep{Rajaguru2006}. Tested by numerical simulations \citep{Parchevsky2008,Hanasoge2008}.}
 \vspace*{1mm}\\
 \multicolumn{1}{p{5.50cm}}{\raggedright Inclined-field ("shower glass") effect \citep{Schunker2005}.} &
 \multicolumn{1}{p{5.50cm}}{\raggedright Travel-time shifts $\approx ten$~seconds. Absent in intensity data \citep{Zhao2006}. Tested by numerical simulations \citep{Parchevsky2009}.}
 \vspace*{1mm}\\
 \multicolumn{1}{p{5.50cm}}{\raggedright Validity of the ray-path theory; finite-wavelength effects \citep{Bogdan1997} } &
 \multicolumn{1}{p{5.50cm}}{\raggedright Tested by using a Born-approximation \citep{Birch2001,Couvidat2006}}
 \vspace*{1mm}\\
 \multicolumn{1}{p{5.50cm}}{\raggedright Differences in travel-time definitions: Gabor wavelet \citep{Kosovichev1997}, minimization of cross-covariance deviation \citep{Gizon2002}, and linearization of the deviation \citep{Gizon2004}} &
 \multicolumn{1}{p{5.50cm}}{\raggedright Tests using MDI data \citep{Couvidat2010} found good agreement between the trave-time definitions of \citet{Kosovichev1997} and \citet{Gizon2002}, but strong systematic deviations of the linearized definition of \citet{Gizon2004}.}
 \vspace*{1mm}\\
 \multicolumn{1}{p{5.50cm}}{\raggedright Contributions of thermal and magnetic effects \citep{Kosovichev1997}
} &
 \multicolumn{1}{p{5.50cm}}{\raggedright  Tested by numerical simulations \citep{Olshevsky2008,Shelyag2009a}. Results are model dependent.}
 \vspace*{1mm}\\
 \multicolumn{1}{p{5.50cm}}{\raggedright Transformation in different types of MHD waves} &
 \multicolumn{1}{p{5.50cm}}{\raggedright Studied by numerical simulations \citep{Parchevsky2009,Parchevsky2010} and wave-form analysis of observations \citep{Zhao2010a}. No significant effect has been found.}
 \\
 \vspace*{2mm}\\
 \multicolumn{1}{p{5.50cm}}{\raggedright Relationship between surface moat outflow and deep flows} &
 \multicolumn{1}{p{5.50cm}}{\raggedright Moat outflow is observed in {\it f}-mode travel times \citep{Gizon2001a}; the {\it p}-mode travel times correspond to deep inflows \citep{Zhao2001,Zhao2010}. A unified inversion procedure for {\it f}- and {\it p}-mode has not been developed.}
 \\
 \vspace*{2mm}\\
 \multicolumn{1}{p{5.50cm}}{\raggedright Cross-talk between the horizontal and vertical velocities} &
 \multicolumn{1}{p{5.50cm}}{\raggedright Studied by modeling \citep{Zhao2003a}. It can result in underestimation of downward velocity beneath sunspots, but no artificial sign reversal. Improvement of the inversion procedure is needed \citep{Jackiewicz2009}.}
 \\
 \vspace*{2mm}\\
 \multicolumn{1}{p{5.50cm}}{\raggedright Comparison of the time-distance and ring-diagram results} &
 \multicolumn{1}{p{5.50cm}}{\raggedright The results are in a general qualitative agreement \citep{Hindman2003,Hindman2004,Basu2004,Bogart2008}. More systematic studies are needed for quantitative comparison because of the large difference in the spatial and temporal resolutions.}
 \\
\hline
\end{tabular}
\end{table}

Many concerns about the local helioseismology inferences have been
resolved by data analysis and numerical simulations. An important
role is played by analysis of observational data from various
sources.

For instance, the helioseismic observations from the {\it Hinode}
resolved several concerns about the reliability of the MDI
measurements. In particular, one concern was about the influence of
the ``inclined-field or magnetic shower-glass effect''
\citep{Schunker2005,Schunker2007}, which shows that helioseismic
travel times measured from Dopplergrams may depend on the
line-of-sight angle in inclined magnetic fields of active regions.
However, the {\it Hinode} results were obtained from the Ca~{\sc ii}~H intensity
data, and this effect does not exist in the travel times measured
from intensity oscillations \citep{Zhao2006}. Therefore, the Hinode data
confirmed that the previous results were not significantly affected
by the inclined field "shower glass" effect. The {\it Hinode} data have
also allowed us to qualitatively confirm the MDI time--distance helioseismology
results, previously
obtained with phase-speed filtering. The phase-speed filtering
procedure substantially improves the signal-to-noise ratio in the travel-time
measurements and will remain an essential component of this technique.
Thus, it is important to take into account this procedure in the
travel-time sensitivity kernels calculated using the Born approximation
\citep{Gizon2002,Birch2004a}.

In addition, the important problems of local-helioseismology inversions that need to be resolved
are the separation of magnetic and thermal effects, development of a unified
procedure for inversion of {\it f}- and {\it p}-mode travel times, and improvement
of inferences of the vertical flow component in both time--distance and ring-diagram
techniques. The magnetized subsurface turbulence certainly plays a very important role in the
oscillation physics, and these effects must be investigated.

Numerical simulations become increasingly important in investigations of the complicated
physics of wave excitation, propagation, and interaction with magnetic regions. The synergy
between the simulations and observation will allow us to improve helioseismology
techniques and understanding of the sunspot structure and dynamics. The uninterrupted high-resolution helioseismology data from  {\it Solar Dynamics Observatory} provide new opportunities for detailed investigation of the process of emergence of magnetic flux, formation, and evolution of sunspots.

\newpage
\bibliographystyle{spr-mp-sola-cnd}


\begin{thebibliography}{110}
\ifx \bisbn   \undefined \def \bisbn  #1{ISBN #1}\fi
\ifx \binits  \undefined \def \binits#1{#1} \fi
\ifx \bauthor  \undefined \def \bauthor#1{#1} \fi
\ifx \batitle  \undefined \def \batitle#1{#1} \fi
\ifx \bjtitle  \undefined \def \bjtitle#1{\textit{#1}}\fi
\ifx \bvolume  \undefined \def \bvolume#1{\textbf{#1}}\fi
\ifx \byear  \undefined \def \byear#1{#1} \fi
\ifx \bissue  \undefined \def \bissue#1{#1} \fi
\ifx \bfpage  \undefined \def \bfpage#1{#1} \fi
\ifx \blpage  \undefined \def \blpage #1{#1} \fi
\ifx \burl  \undefined \def \burl#1{\textsf{#1}} \fi
\ifx \href  \undefined \def \href#1#2{#2} \fi
\ifx \doiurl  \undefined \def \doiurl#1{\href{http://dx.doi.org/#1}{#1}} \fi
\ifx \betal  \undefined \def \betal{\textit{et al.}} \fi
\ifx \binstitute  \undefined \def \binstitute#1{#1} \fi
\ifx \bctitle  \undefined \def \bctitle#1{#1} \fi
\ifx \beditor  \undefined \def \beditor#1{#1} \fi
\ifx \bpublisher  \undefined \def \bpublisher#1{#1} \fi
\ifx \bbtitle  \undefined \def \bbtitle#1{\textit{#1}} \fi
\ifx \bedition  \undefined \def \bedition#1{#1} \fi
\ifx \bseriesno  \undefined \def \bseriesno#1{\textbf{#1}} \fi
\ifx \blocation  \undefined \def \blocation#1{#1} \fi
\ifx \bsertitle  \undefined \def \bsertitle#1{\textit{#1}} \fi
\ifx \bsnm \undefined \def \bsnm#1{#1} \fi
\ifx \bsuffix \undefined \def \bsuffix#1{#1} \fi
\ifx \bparticle \undefined \def \bparticle#1{#1} \fi
\ifx \barticle \undefined \def \barticle#1{#1} \fi
\ifx \botherref \undefined \def \botherref #1{#1} \fi
\ifx \url \undefined \def \url#1{\textsf{#1}} \fi
\ifx \bchapter \undefined \def \bchapter#1{#1} \fi
\ifx \bbook \undefined \def \bbook#1{#1} \fi
\ifx \bcomment \undefined \def \bcomment#1{#1} \fi
\ifx \oauthor \undefined \def \oauthor#1{#1} \fi
\ifx \citeauthoryear \undefined \def \citeauthoryear#1{#1} \fi
\def \endbibitem {}

\bibitem[\protect\citeauthoryear{{Baldner} \textit{et~al.}}{2009}]{Baldner2009}
\begin{botherref}
\oauthor{\bsnm{{Baldner}}, \binits{C.S.}}, \oauthor{\bsnm{{Bogart}},
  \binits{R.S.}}, \oauthor{\bsnm{{Basu}}, \binits{S.}},
  \oauthor{\bsnm{{Antia}}, \binits{H.M.}}:
2009,
In: {Dikpati~M., Arentoft~T., Gonz{\'a}lez Hern{\'a}ndez,~I, Lindsey,~C. \&
  Hill,~F.} (eds)
\textit{Solar-Stellar Dynamos as Revealed by Helio- and Asteroseismology: GONG 2008/SOHO 21},
Astron. Soc. Pac., San Francisco,
\textbf{416},
119.
\end{botherref}
\endbibitem

\bibitem[\protect\citeauthoryear{{Balthasar} and
  {Muglach}}{2010}]{Balthasar2010}
\begin{barticle}
\bauthor{\bsnm{{Balthasar}}, \binits{H.}}, \bauthor{\bsnm{{Muglach}},
  \binits{K.}}:
\byear{2010},
\bjtitle{\aap}
\bvolume{511},
\bfpage{A67}.
\url{doi:10.1051/0004-6361/200912978}.
\end{barticle}
\endbibitem

\bibitem[\protect\citeauthoryear{{Basu}, {Antia}, and
  {Bogart}}{2004}]{Basu2004}
\begin{barticle}
\bauthor{\bsnm{{Basu}}, \binits{S.}}, \bauthor{\bsnm{{Antia}}, \binits{H.M.}},
  \bauthor{\bsnm{{Bogart}}, \binits{R.S.}}:
\byear{2004},
\bjtitle{\apj}
\bvolume{610},
\bfpage{1157}.
\url{doi:10.1086/421843}.
\end{barticle}
\endbibitem

\bibitem[\protect\citeauthoryear{{Birch} and {Kosovichev}}{2000}]{Birch2000}
\begin{barticle}
\bauthor{\bsnm{{Birch}}, \binits{A.C.}}, \bauthor{\bsnm{{Kosovichev}},
  \binits{A.G.}}:
\byear{2000},
\bjtitle{\solphys}
\bvolume{192},
\bfpage{193}.
\end{barticle}
\endbibitem

\bibitem[\protect\citeauthoryear{{Birch}, {Kosovichev}, and
  {Duvall}}{2004}]{Birch2004a}
\begin{barticle}
\bauthor{\bsnm{{Birch}}, \binits{A.C.}}, \bauthor{\bsnm{{Kosovichev}},
  \binits{A.G.}}, \bauthor{\bsnm{{Duvall}}, \binits{T.L.} \bsuffix{Jr.}}:
\byear{2004},
\bjtitle{\apj}
\bvolume{608},
\bfpage{580}.
\url{doi:10.1086/386361}.
\end{barticle}
\endbibitem

\bibitem[\protect\citeauthoryear{{Birch} \textit{et~al.}}{2001}]{Birch2001}
\begin{barticle}
\bauthor{\bsnm{{Birch}}, \binits{A.C.}}, \bauthor{\bsnm{{Kosovichev}},
  \binits{A.G.}}, \bauthor{\bsnm{{Price}}, \binits{G.H.}},
  \bauthor{\bsnm{{Schlottmann}}, \binits{R.B.}}:
\byear{2001},
\bjtitle{\apjl}
\bvolume{561},
\bfpage{L229}.
\url{doi:10.1086/324766}.
\end{barticle}
\endbibitem

\bibitem[\protect\citeauthoryear{{Birch} \textit{et~al.}}{2009}]{Birch2009}
\begin{barticle}
\bauthor{\bsnm{{Birch}}, \binits{A.C.}}, \bauthor{\bsnm{{Braun}},
  \binits{D.C.}}, \bauthor{\bsnm{{Hanasoge}}, \binits{S.M.}},
  \bauthor{\bsnm{{Cameron}}, \binits{R.}}:
\byear{2009},
\bjtitle{\solphys}
\bvolume{254},
\bfpage{17}.
\url{doi:10.1007/s11207-008-9282-9}.
\end{barticle}
\endbibitem

\bibitem[\protect\citeauthoryear{{Bogart} \textit{et~al.}}{2008}]{Bogart2008}
\begin{barticle}
\bauthor{\bsnm{{Bogart}}, \binits{R.S.}}, \bauthor{\bsnm{{Basu}}, \binits{S.}},
  \bauthor{\bsnm{{Rabello-Soares}}, \binits{M.C.}}, \bauthor{\bsnm{{Antia}},
  \binits{H.M.}}:
\byear{2008},
\bjtitle{\solphys}
\bvolume{251},
\bfpage{439}.
\url{doi:10.1007/s11207-008-9213-9}.
\end{barticle}
\endbibitem

\bibitem[\protect\citeauthoryear{{Bogdan}}{1997}]{Bogdan1997}
\begin{barticle}
\bauthor{\bsnm{{Bogdan}}, \binits{T.J.}}:
\byear{1997},
\bjtitle{\apj}
\bvolume{477},
\bfpage{475}.
\url{doi:10.1086/303680}.
\end{barticle}
\endbibitem

\bibitem[\protect\citeauthoryear{{Botha}, {Rucklidge}, and
  {Hurlburt}}{2006}]{Botha2006}
\begin{barticle}
\bauthor{\bsnm{{Botha}}, \binits{G.J.J.}}, \bauthor{\bsnm{{Rucklidge}},
  \binits{A.M.}}, \bauthor{\bsnm{{Hurlburt}}, \binits{N.E.}}:
\byear{2006},
\bjtitle{\mnras}
\bvolume{369},
\bfpage{1611}.
\url{doi:10.1111/j.1365-2966.2006.10480.x}.
\end{barticle}
\endbibitem

\bibitem[\protect\citeauthoryear{{Botha}, {Rucklidge}, and
  {Hurlburt}}{2007}]{Botha2007}
\begin{barticle}
\bauthor{\bsnm{{Botha}}, \binits{G.J.J.}}, \bauthor{\bsnm{{Rucklidge}},
  \binits{A.M.}}, \bauthor{\bsnm{{Hurlburt}}, \binits{N.E.}}:
\byear{2007},
\bjtitle{\apjl}
\bvolume{662},
\bfpage{L27}.
\url{doi:10.1086/519079}.
\end{barticle}
\endbibitem

\bibitem[\protect\citeauthoryear{{Botha} \textit{et~al.}}{2008}]{Botha2008}
\begin{barticle}
\bauthor{\bsnm{{Botha}}, \binits{G.J.J.}}, \bauthor{\bsnm{{Busse}},
  \binits{F.H.}}, \bauthor{\bsnm{{Hurlburt}}, \binits{N.E.}},
  \bauthor{\bsnm{{Rucklidge}}, \binits{A.M.}}:
\byear{2008},
\bjtitle{\mnras}
\bvolume{387},
\bfpage{1445}.
\url{doi:10.1111/j.1365-2966.2008.13359.x}.
\end{barticle}
\endbibitem

\bibitem[\protect\citeauthoryear{{Cally}}{2009}]{Cally2009}
\begin{barticle}
\bauthor{\bsnm{{Cally}}, \binits{P.S.}}:
\byear{2009},
\bjtitle{\mnras}
\bvolume{395},
\bfpage{1309}.
\url{doi:10.1111/j.1365-2966.2009.14708.x}.
\end{barticle}
\endbibitem

\bibitem[\protect\citeauthoryear{{Cameron} \textit{et~al.}}{2010}]{Cameron2010}
\begin{botherref}
\oauthor{\bsnm{{Cameron}}, \binits{R.}}, \oauthor{\bsnm{{Gizon}}, \binits{L.}},
  \oauthor{\bsnm{{Schunker}}, \binits{H.}}, \oauthor{\bsnm{{Pietarila}},
  \binits{A.}}:
2010,
\textit{\solphys}, in press.
\end{botherref}
\endbibitem

\bibitem[\protect\citeauthoryear{{Couvidat}, {Birch}, and
  {Kosovichev}}{2006{\natexlab{a}}}]{Couvidat2006}
\begin{barticle}
\bauthor{\bsnm{{Couvidat}}, \binits{S.}}, \bauthor{\bsnm{{Birch}},
  \binits{A.C.}}, \bauthor{\bsnm{{Kosovichev}}, \binits{A.G.}}:
\byear{2006},
\bjtitle{\apj}
\bvolume{640},
\bfpage{516}.
\url{doi:10.1086/500103}.
\end{barticle}
\endbibitem

\bibitem[\protect\citeauthoryear{{Couvidat}, {Birch}, and
  {Kosovichev}}{2006{\natexlab{b}}}]{Couvidat2006c}
\begin{barticle}
\bauthor{\bsnm{{Couvidat}}, \binits{S.}}, \bauthor{\bsnm{{Birch}},
  \binits{A.C.}}, \bauthor{\bsnm{{Kosovichev}}, \binits{A.G.}}:
\byear{2006},
\bjtitle{\apj}
\bvolume{640},
\bfpage{516}.
\url{doi:10.1086/500103}.
\end{barticle}
\endbibitem

\bibitem[\protect\citeauthoryear{{Couvidat}
  \textit{et~al.}}{2004}]{Couvidat2004}
\begin{barticle}
\bauthor{\bsnm{{Couvidat}}, \binits{S.}}, \bauthor{\bsnm{{Birch}},
  \binits{A.C.}}, \bauthor{\bsnm{{Kosovichev}}, \binits{A.G.}},
  \bauthor{\bsnm{{Zhao}}, \binits{J.}}:
\byear{2004},
\bjtitle{\apj}
\bvolume{607},
\bfpage{554}.
\url{doi:10.1086/383342}.
\end{barticle}
\endbibitem

\bibitem[\protect\citeauthoryear{{Couvidat}
  \textit{et~al.}}{2006{\natexlab{c}}}]{Couvidat2006a}
\begin{botherref}
\oauthor{\bsnm{{Couvidat}}, \binits{S.}}, \oauthor{\bsnm{{Birch}},
  \binits{A.C.}}, \oauthor{\bsnm{{Rajaguru}}, \binits{S.P.}},
  \oauthor{\bsnm{{Kosovichev}}, \binits{A.G.}}:
2006,
In: {Bothmer~V. \& Hady~A.~A.} (eds)
\textit{Solar Activity and its Magnetic Origin},
\textit{IAU Symposium}, Cambridge Univ. Press, Cambridge,
\textbf{233},
75.
\url{doi:10.1017/S1743921306001499}.
\end{botherref}
\endbibitem

\bibitem[\protect\citeauthoryear{{Couvidat}
  \textit{et~al.}}{2010}]{Couvidat2010}
\begin{botherref}
\oauthor{\bsnm{{Couvidat}}, \binits{S.}}, \oauthor{\bsnm{{Zhao}}, \binits{J.}},
  \oauthor{\bsnm{{Birch}}, \binits{A.C.}}, \oauthor{\bsnm{{Kosovichev}},
  \binits{A.G.}}, \oauthor{\bsnm{{Duvall, Jr.}}, \binits{T.L.}},
  \oauthor{\bsnm{{Parchevsky}}, \binits{K.V.}}, \oauthor{\bsnm{{Scherrer}},
  \binits{P.H.}}:
2010,
\textit{\solphys}
\textrm{in press}.
\end{botherref}
\endbibitem

\bibitem[\protect\citeauthoryear{{Cowling}}{1946}]{Cowling1946}
\begin{barticle}
\bauthor{\bsnm{{Cowling}}, \binits{T.G.}}:
\byear{1946},
\bjtitle{\mnras}
\bvolume{106},
\bfpage{218}.
\end{barticle}
\endbibitem

\bibitem[\protect\citeauthoryear{{Duvall} \textit{et~al.}}{1993}]{Duvall1993}
\begin{barticle}
\bauthor{\bsnm{{Duvall}}, \binits{T.L.} \bsuffix{Jr.}},
  \bauthor{\bsnm{{Jefferies}}, \binits{S.M.}}, \bauthor{\bsnm{{Harvey}},
  \binits{J.W.}}, \bauthor{\bsnm{{Pomerantz}}, \binits{M.A.}}:
\byear{1993},
\bjtitle{\nat}
\bvolume{362},
\bfpage{430}.
\url{doi:10.1038/362430a0}.
\end{barticle}
\endbibitem

\bibitem[\protect\citeauthoryear{{Duvall} \textit{et~al.}}{1997}]{Duvall1997}
\begin{barticle}
\bauthor{\bsnm{{Duvall}}, \binits{T.L.} \bsuffix{Jr.}},
  \bauthor{\bsnm{{Kosovichev}}, \binits{A.G.}}, \bauthor{\bsnm{{Scherrer}},
  \binits{P.H.}}, \bauthor{\bsnm{{Bogart}}, \binits{R.S.}},
  \bauthor{\bsnm{{Bush}}, \binits{R.I.}}, \bauthor{\bsnm{{de Forest}},
  \binits{C.}}, \bauthor{\bsnm{{Hoeksema}}, \binits{J.T.}},
  \bauthor{\bsnm{{Schou}}, \binits{J.}}, \bauthor{\bsnm{{Saba}},
  \binits{J.L.R.}}, \bauthor{\bsnm{{Tarbell}}, \binits{T.D.}},
  \bauthor{\bsnm{{Title}}, \binits{A.M.}}, \bauthor{\bsnm{{Wolfson}},
  \binits{C.J.}}, \bauthor{\bsnm{{Milford}}, \binits{P.N.}}:
\byear{1997},
\bjtitle{\solphys}
\bvolume{170},
\bfpage{63}.
\end{barticle}
\endbibitem

\bibitem[\protect\citeauthoryear{{Duvall} \textit{et~al.}}{1996}]{Duvall1996}
\begin{barticle}
\bauthor{\bsnm{{Duvall}}, \binits{T.L.}}, \bauthor{\bsnm{{D'Silva}},
  \binits{S.}}, \bauthor{\bsnm{{Jefferies}}, \binits{S.M.}},
  \bauthor{\bsnm{{Harvey}}, \binits{J.W.}}, \bauthor{\bsnm{{Schou}},
  \binits{J.}}:
\byear{1996},
\bjtitle{\nat}
\bvolume{379},
\bfpage{235}.
\url{doi:10.1038/379235a0}.
\end{barticle}
\endbibitem

\bibitem[\protect\citeauthoryear{{Evershed}}{1909}]{Evershed1909}
\begin{barticle}
\bauthor{\bsnm{{Evershed}}, \binits{J.}}:
\byear{1909},
\bjtitle{\mnras}
\bvolume{69},
\bfpage{454}.
\end{barticle}
\endbibitem

\bibitem[\protect\citeauthoryear{{Gizon} and {Birch}}{2002}]{Gizon2002}
\begin{barticle}
\bauthor{\bsnm{{Gizon}}, \binits{L.}}, \bauthor{\bsnm{{Birch}}, \binits{A.C.}}:
\byear{2002},
\bjtitle{\apj}
\bvolume{571},
\bfpage{966}.
\url{doi:10.1086/340015}.
\end{barticle}
\endbibitem

\bibitem[\protect\citeauthoryear{{Gizon} and {Birch}}{2004}]{Gizon2004}
\begin{barticle}
\bauthor{\bsnm{{Gizon}}, \binits{L.}}, \bauthor{\bsnm{{Birch}}, \binits{A.C.}}:
\byear{2004},
\bjtitle{\apj}
\bvolume{614},
\bfpage{472}.
\url{doi:10.1086/423367}.
\end{barticle}
\endbibitem

\bibitem[\protect\citeauthoryear{{Gizon}, {Duvall}, and
  {Larsen}}{2001{\natexlab{a}}}]{Gizon2001a}
\begin{botherref}
\oauthor{\bsnm{{Gizon}}, \binits{L.}}, \oauthor{\bsnm{{Duvall}}, \binits{T.L.}
  \bsuffix{Jr.}}, \oauthor{\bsnm{{Larsen}}, \binits{R.M.}}:
2001,
In: {Brekke~P., Fleck~B., \& Gurman~J.~B.} (eds)
\textit{Recent Insights into the Physics of the Sun and Heliosphere: Highlights
  from SOHO and Other Space Missions},
\textit{IAU Symposium}
\textbf{203}, Cambridge Univ. Press, Cambridge,
189.
\end{botherref}
\endbibitem

\bibitem[\protect\citeauthoryear{{Gizon}
  \textit{et~al.}}{2001{\natexlab{b}}}]{Gizon2001}
\begin{botherref}
\oauthor{\bsnm{{Gizon}}, \binits{L.}}, \oauthor{\bsnm{{Birch}}, \binits{A.C.}},
  \oauthor{\bsnm{{Bush}}, \binits{R.I.}}, \oauthor{\bsnm{{Duvall}},
  \binits{T.L.} \bsuffix{Jr.}}, \oauthor{\bsnm{{Kosovichev}}, \binits{A.G.}},
  \oauthor{\bsnm{{Scherrer}}, \binits{P.H.}}, \oauthor{\bsnm{{Zhao}},
  \binits{J.}}:
2001,
In: {Battrick~B., Sawaya-Lacoste~H., Marsch~E., Martinez Pillet~V., Fleck~B.,
  \& Marsden~R.} (eds)
\textit{Solar encounter. Proceedings of the First Solar Orbiter Workshop},
SP-\textbf{493}, ESA, Noordwijk,
227.
\end{botherref}
\endbibitem

\bibitem[\protect\citeauthoryear{{Gizon} \textit{et~al.}}{2009}]{Gizon2009}
\begin{barticle}
\bauthor{\bsnm{{Gizon}}, \binits{L.}}, \bauthor{\bsnm{{Schunker}},
  \binits{H.}}, \bauthor{\bsnm{{Baldner}}, \binits{C.S.}},
  \bauthor{\bsnm{{Basu}}, \binits{S.}}, \bauthor{\bsnm{{Birch}},
  \binits{A.C.}}, \bauthor{\bsnm{{Bogart}}, \binits{R.S.}},
  \bauthor{\bsnm{{Braun}}, \binits{D.C.}}, \bauthor{\bsnm{{Cameron}},
  \binits{R.}}, \bauthor{\bsnm{{Duvall}}, \binits{T.L.}},
  \bauthor{\bsnm{{Hanasoge}}, \binits{S.M.}}, \bauthor{\bsnm{{Jackiewicz}},
  \binits{J.}}, \bauthor{\bsnm{{Roth}}, \binits{M.}}, \bauthor{\bsnm{{Stahn}},
  \binits{T.}}, \bauthor{\bsnm{{Thompson}}, \binits{M.J.}},
  \bauthor{\bsnm{{Zharkov}}, \binits{S.}}:
\byear{2009},
\bjtitle{Space Science Rev.}
\bvolume{144},
\bfpage{249}.
\url{doi:10.1007/s11214-008-9466-5}.
\end{barticle}
\endbibitem

\bibitem[\protect\citeauthoryear{{Gough} and {Toomre}}{1983}]{Gough1983}
\begin{barticle}
\bauthor{\bsnm{{Gough}}, \binits{D.O.}}, \bauthor{\bsnm{{Toomre}},
  \binits{J.}}:
\byear{1983},
\bjtitle{\solphys}
\bvolume{82},
\bfpage{401}.
\end{barticle}
\endbibitem

\bibitem[\protect\citeauthoryear{{Haber} \textit{et~al.}}{2000}]{Haber2000}
\begin{barticle}
\bauthor{\bsnm{{Haber}}, \binits{D.A.}}, \bauthor{\bsnm{{Hindman}},
  \binits{B.W.}}, \bauthor{\bsnm{{Toomre}}, \binits{J.}},
  \bauthor{\bsnm{{Bogart}}, \binits{R.S.}}, \bauthor{\bsnm{{Thompson}},
  \binits{M.J.}}, \bauthor{\bsnm{{Hill}}, \binits{F.}}:
\byear{2000},
\bjtitle{\solphys}
\bvolume{192},
\bfpage{335}.
\end{barticle}
\endbibitem

\bibitem[\protect\citeauthoryear{{Haber} \textit{et~al.}}{2002}]{Haber2002}
\begin{barticle}
\bauthor{\bsnm{{Haber}}, \binits{D.A.}}, \bauthor{\bsnm{{Hindman}},
  \binits{B.W.}}, \bauthor{\bsnm{{Toomre}}, \binits{J.}},
  \bauthor{\bsnm{{Bogart}}, \binits{R.S.}}, \bauthor{\bsnm{{Larsen}},
  \binits{R.M.}}, \bauthor{\bsnm{{Hill}}, \binits{F.}}:
\byear{2002},
\bjtitle{\apj}
\bvolume{570},
\bfpage{855}.
\url{doi:10.1086/339631}.
\end{barticle}
\endbibitem

\bibitem[\protect\citeauthoryear{{Haber} \textit{et~al.}}{2004}]{Haber2004}
\begin{barticle}
\bauthor{\bsnm{{Haber}}, \binits{D.A.}}, \bauthor{\bsnm{{Hindman}},
  \binits{B.W.}}, \bauthor{\bsnm{{Toomre}}, \binits{J.}},
  \bauthor{\bsnm{{Thompson}}, \binits{M.J.}}:
\byear{2004},
\bjtitle{\solphys}
\bvolume{220},
\bfpage{371}.
\url{doi:10.1023/B:SOLA.0000031405.52911.08}.
\end{barticle}
\endbibitem

\bibitem[\protect\citeauthoryear{{Hanasoge}
  \textit{et~al.}}{2008}]{Hanasoge2008}
\begin{barticle}
\bauthor{\bsnm{{Hanasoge}}, \binits{S.M.}}, \bauthor{\bsnm{{Couvidat}},
  \binits{S.}}, \bauthor{\bsnm{{Rajaguru}}, \binits{S.P.}},
  \bauthor{\bsnm{{Birch}}, \binits{A.C.}}:
\byear{2008},
\bjtitle{\mnras}
\bvolume{391},
\bfpage{1931}.
\url{doi:10.1111/j.1365-2966.2008.14013.x}.
\end{barticle}
\endbibitem

\bibitem[\protect\citeauthoryear{{Hartlep} \textit{et~al.}}{2010}]{Hartlep2010}
\begin{botherref}
\oauthor{\bsnm{{Hartlep}}, \binits{T.}}, \oauthor{\bsnm{{Busse}},
  \binits{F.H.}}, \oauthor{\bsnm{{Hulburt}}, \binits{N.E.}},
  \oauthor{\bsnm{{Kosovichev}}, \binits{A.G.}}:
2010,
\textit{ArXiv e-prints}
\textrm{1006.4156}.
\end{botherref}
\endbibitem

\bibitem[\protect\citeauthoryear{{Harvey} and {Harvey}}{1973}]{Harvey1973}
\begin{barticle}
\bauthor{\bsnm{{Harvey}}, \binits{K.}}, \bauthor{\bsnm{{Harvey}}, \binits{J.}}:
\byear{1973},
\bjtitle{\solphys}
\bvolume{28},
\bfpage{61}.
\url{doi:10.1007/BF00152912}.
\end{barticle}
\endbibitem

\bibitem[\protect\citeauthoryear{{Heinemann}
  \textit{et~al.}}{2007}]{Heinemann2007}
\begin{barticle}
\bauthor{\bsnm{{Heinemann}}, \binits{T.}}, \bauthor{\bsnm{{Nordlund}},
  \binits{{\AA}.}}, \bauthor{\bsnm{{Scharmer}}, \binits{G.B.}},
  \bauthor{\bsnm{{Spruit}}, \binits{H.C.}}:
\byear{2007},
\bjtitle{\apj}
\bvolume{669},
\bfpage{1390}.
\url{doi:10.1086/520827}.
\end{barticle}
\endbibitem

\bibitem[\protect\citeauthoryear{{Hill}}{1988}]{Hill1988}
\begin{barticle}
\bauthor{\bsnm{{Hill}}, \binits{F.}}:
\byear{1988},
\bjtitle{\apj}
\bvolume{333},
\bfpage{996}.
\url{doi:10.1086/166807}.
\end{barticle}
\endbibitem

\bibitem[\protect\citeauthoryear{{Hindman}, {Haber}, and
  {Toomre}}{2006}]{Hindman2006}
\begin{barticle}
\bauthor{\bsnm{{Hindman}}, \binits{B.W.}}, \bauthor{\bsnm{{Haber}},
  \binits{D.A.}}, \bauthor{\bsnm{{Toomre}}, \binits{J.}}:
\byear{2006},
\bjtitle{\apj}
\bvolume{653},
\bfpage{725}.
\url{doi:10.1086/508603}.
\end{barticle}
\endbibitem

\bibitem[\protect\citeauthoryear{{Hindman}, {Haber}, and
  {Toomre}}{2009}]{Hindman2009}
\begin{barticle}
\bauthor{\bsnm{{Hindman}}, \binits{B.W.}}, \bauthor{\bsnm{{Haber}},
  \binits{D.A.}}, \bauthor{\bsnm{{Toomre}}, \binits{J.}}:
\byear{2009},
\bjtitle{\apj}
\bvolume{698},
\bfpage{1749}.
\url{doi:10.1088/0004-637X/698/2/1749}.
\end{barticle}
\endbibitem

\bibitem[\protect\citeauthoryear{{Hindman} \textit{et~al.}}{2003}]{Hindman2003}
\begin{botherref}
\oauthor{\bsnm{{Hindman}}, \binits{B.W.}}, \oauthor{\bsnm{{Zhao}},
  \binits{J.}}, \oauthor{\bsnm{{Haber}}, \binits{D.A.}},
  \oauthor{\bsnm{{Kosovichev}}, \binits{A.G.}}, \oauthor{\bsnm{{Toomre}},
  \binits{J.}}:
2003,
\textit{Bull. Am. Astronom. Soc.},
\textbf{35},
822.
\end{botherref}
\endbibitem

\bibitem[\protect\citeauthoryear{{Hindman} \textit{et~al.}}{2004}]{Hindman2004}
\begin{barticle}
\bauthor{\bsnm{{Hindman}}, \binits{B.W.}}, \bauthor{\bsnm{{Gizon}},
  \binits{L.}}, \bauthor{\bsnm{{Duvall}}, \binits{T.L.} \bsuffix{Jr.}},
  \bauthor{\bsnm{{Haber}}, \binits{D.A.}}, \bauthor{\bsnm{{Toomre}},
  \binits{J.}}:
\byear{2004},
\bjtitle{\apj}
\bvolume{613},
\bfpage{1253}.
\url{doi:10.1086/423263}.
\end{barticle}
\endbibitem

\bibitem[\protect\citeauthoryear{{Howe}}{2008}]{Howe2008}
\begin{barticle}
\bauthor{\bsnm{{Howe}}, \binits{R.}}:
\byear{2008},
\bjtitle{Adv. Space Res.}
\bvolume{41},
\bfpage{846}.
\url{doi:10.1016/j.asr.2006.12.033}.
\end{barticle}
\endbibitem

\bibitem[\protect\citeauthoryear{{Hurlburt} and {De Rosa}}{2008}]{Hurlburt2008}
\begin{barticle}
\bauthor{\bsnm{{Hurlburt}}, \binits{N.}}, \bauthor{\bsnm{{De Rosa}},
  \binits{M.}}:
\byear{2008},
\bjtitle{\apjl}
\bvolume{684},
\bfpage{L123}.
\url{doi:10.1086/591736}.
\end{barticle}
\endbibitem

\bibitem[\protect\citeauthoryear{{Hurlburt} and
  {Rucklidge}}{2000}]{Hurlburt2000}
\begin{barticle}
\bauthor{\bsnm{{Hurlburt}}, \binits{N.E.}}, \bauthor{\bsnm{{Rucklidge}},
  \binits{A.M.}}:
\byear{2000},
\bjtitle{\mnras}
\bvolume{314},
\bfpage{793}.
\url{doi:10.1046/j.1365-8711.2000.03407.x}.
\end{barticle}
\endbibitem

\bibitem[\protect\citeauthoryear{{Jackiewicz}}{2009}]{Jackiewicz2009}
\begin{botherref}
\oauthor{\bsnm{{Jackiewicz}}, \binits{J.}}:
2009,
In: {Guzik~J.~A. \& Bradley~P.~A.} (eds)
\textit{Stellar Pulsation: Challenges for Theory and Observation},
\textit{Am. Inst. Phys. Conf. Ser.},
\textbf{1170},
574.
\url{doi:10.1063/1.3246565}.
\end{botherref}
\endbibitem

\bibitem[\protect\citeauthoryear{{Jahn}}{1992}]{Jahn1992}
\begin{botherref}
\oauthor{\bsnm{{Jahn}}, \binits{K.}}:
1992,
In: {Thomas~J.~H. \& Weiss~N.~O.} (eds)
\textit{Sunspots. Theory and Observations}, \textit{Proc. NATO Adv. Res. Workshop, Cambridge, UK}, Kluwer Acad. Publ.,
139.
\end{botherref}
\endbibitem

\bibitem[\protect\citeauthoryear{{Jensen} \textit{et~al.}}{2001}]{Jensen2001}
\begin{barticle}
\bauthor{\bsnm{{Jensen}}, \binits{J.M.}}, \bauthor{\bsnm{{Duvall}},
  \binits{T.L.} \bsuffix{Jr.}}, \bauthor{\bsnm{{Jacobsen}}, \binits{B.H.}},
  \bauthor{\bsnm{{Christensen-Dalsgaard}}, \binits{J.}}:
\byear{2001},
\bjtitle{\apjl}
\bvolume{553},
\bfpage{L193}.
\url{doi:10.1086/320677}.
\end{barticle}
\endbibitem

\bibitem[\protect\citeauthoryear{{Khomenko} and
  {Collados}}{2008}]{Khomenko2008}
\begin{barticle}
\bauthor{\bsnm{{Khomenko}}, \binits{E.}}, \bauthor{\bsnm{{Collados}},
  \binits{M.}}:
\byear{2008},
\bjtitle{\apj}
\bvolume{689},
\bfpage{1379}.
\url{doi:10.1086/592681}.
\end{barticle}
\endbibitem

\bibitem[\protect\citeauthoryear{{Khomenko}
  \textit{et~al.}}{2009}]{Khomenko2009}
\begin{barticle}
\bauthor{\bsnm{{Khomenko}}, \binits{E.}}, \bauthor{\bsnm{{Kosovichev}},
  \binits{A.}}, \bauthor{\bsnm{{Collados}}, \binits{M.}},
  \bauthor{\bsnm{{Parchevsky}}, \binits{K.}}, \bauthor{\bsnm{{Olshevsky}},
  \binits{V.}}:
\byear{2009},
\bjtitle{\apj}
\bvolume{694},
\bfpage{411}.
\url{doi:10.1088/0004-637X/694/1/411}.
\end{barticle}
\endbibitem

\bibitem[\protect\citeauthoryear{{Kitiashvili}
  \textit{et~al.}}{2009}]{Kitiashvili2009}
\begin{barticle}
\bauthor{\bsnm{{Kitiashvili}}, \binits{I.N.}}, \bauthor{\bsnm{{Kosovichev}},
  \binits{A.G.}}, \bauthor{\bsnm{{Wray}}, \binits{A.A.}},
  \bauthor{\bsnm{{Mansour}}, \binits{N.N.}}:
\byear{2009},
\bjtitle{\apjl}
\bvolume{700},
\bfpage{L178}.
\url{doi:10.1088/0004-637X/700/2/L178}.
\end{barticle}
\endbibitem

\bibitem[\protect\citeauthoryear{{Kitiashvili}
  \textit{et~al.}}{2010{\natexlab{a}}}]{Kitiashvili2010}
\begin{barticle}
\bauthor{\bsnm{{Kitiashvili}}, \binits{I.N.}}, \bauthor{\bsnm{{Bellot Rubio}},
  \binits{L.R.}}, \bauthor{\bsnm{{Kosovichev}}, \binits{A.G.}},
  \bauthor{\bsnm{{Mansour}}, \binits{N.N.}}, \bauthor{\bsnm{{Sainz Dalda}},
  \binits{A.}}, \bauthor{\bsnm{{Wray}}, \binits{A.A.}}:
\byear{2010},
\bjtitle{\apjl}
\bvolume{716},
\bfpage{L181}.
\url{doi:10.1088/2041-8205/716/2/L181}.
\end{barticle}
\endbibitem

\bibitem[\protect\citeauthoryear{{Kitiashvili}
  \textit{et~al.}}{2010{\natexlab{b}}}]{Kitiashvili2010a}
\begin{botherref}
\oauthor{\bsnm{{Kitiashvili}}, \binits{I.N.}}, \oauthor{\bsnm{{Kosovichev}},
  \binits{A.G.}}, \oauthor{\bsnm{{Wray}}, \binits{A.A.}},
  \oauthor{\bsnm{{Mansour}}, \binits{N.N.}}:
2010, \textit{\solphys}, in press,
\textit{ArXiv e-prints}
\textrm{1004.2288}.
\end{botherref}
\endbibitem

\bibitem[\protect\citeauthoryear{{Komm}, {Howe}, and
  {Hill}}{2009{\natexlab{a}}}]{Komm2009a}
\begin{barticle}
\bauthor{\bsnm{{Komm}}, \binits{R.}}, \bauthor{\bsnm{{Howe}}, \binits{R.}},
  \bauthor{\bsnm{{Hill}}, \binits{F.}}:
\byear{2009},
\bjtitle{\solphys}
\bvolume{258},
\bfpage{13}.
\url{doi:10.1007/s11207-009-9398-6}.
\end{barticle}
\endbibitem

\bibitem[\protect\citeauthoryear{{Komm}, {Howe}, and
  {Hill}}{2009{\natexlab{b}}}]{Komm2009}
\begin{botherref}
\oauthor{\bsnm{{Komm}}, \binits{R.}}, \oauthor{\bsnm{{Howe}}, \binits{R.}},
  \oauthor{\bsnm{{Hill}}, \binits{F.}}:
2009,
In: {Dikpati~M., Arentoft~T., Gonz{\'a}lez Hern{\'a}ndez,~I, Lindsey,~C. \&
  Hill,~F.} (eds)
\textit{Solar-Stellar Dynamos as Revealed by Helio- and Asteroseismology: GONG 2008/SOHO 21},
Astron. Soc. Pacific, San Francisco,
\textbf{416}, 115.
\end{botherref}
\endbibitem

\bibitem[\protect\citeauthoryear{{Komm} \textit{et~al.}}{2007}]{Komm2007}
\begin{barticle}
\bauthor{\bsnm{{Komm}}, \binits{R.}}, \bauthor{\bsnm{{Howe}}, \binits{R.}},
  \bauthor{\bsnm{{Hill}}, \binits{F.}}, \bauthor{\bsnm{{Miesch}}, \binits{M.}},
  \bauthor{\bsnm{{Haber}}, \binits{D.}}, \bauthor{\bsnm{{Hindman}},
  \binits{B.}}:
\byear{2007},
\bjtitle{\apj}
\bvolume{667},
\bfpage{571}.
\url{doi:10.1086/520765}.
\end{barticle}
\endbibitem

\bibitem[\protect\citeauthoryear{{Komm} \textit{et~al.}}{2008}]{Komm2008}
\begin{barticle}
\bauthor{\bsnm{{Komm}}, \binits{R.}}, \bauthor{\bsnm{{Morita}}, \binits{S.}},
  \bauthor{\bsnm{{Howe}}, \binits{R.}}, \bauthor{\bsnm{{Hill}}, \binits{F.}}:
\byear{2008},
\bjtitle{\apj}
\bvolume{672},
\bfpage{1254}.
\url{doi:10.1086/523998}.
\end{barticle}
\endbibitem

\bibitem[\protect\citeauthoryear{{Kosovichev}}{1996}]{Kosovichev1996}
\begin{barticle}
\bauthor{\bsnm{{Kosovichev}}, \binits{A.G.}}:
\byear{1996},
\bjtitle{\apjl}
\bvolume{461},
\bfpage{L55}.
\url{doi:10.1086/309989}.
\end{barticle}
\endbibitem

\bibitem[\protect\citeauthoryear{{Kosovichev}}{2003}]{Kosovichev2003}
\begin{botherref}
\oauthor{\bsnm{{Kosovichev}}, \binits{A.G.}}:
2003,
In: {Thompson, M.~J.~and Christensen-Dalsgaard, J.} (ed.)
\textit{{Stellar Astrophysical Fluid Dynamics}}, Cambridge Univ. Press, Cambridge,
279.
\end{botherref}
\endbibitem

\bibitem[\protect\citeauthoryear{{Kosovichev}}{2006}]{Kosovichev2006a}
\begin{barticle}
\bauthor{\bsnm{{Kosovichev}}, \binits{A.G.}}:
\byear{2006},
\bjtitle{\solphys}
\bvolume{238},
\bfpage{1}.
\url{doi:10.1007/s11207-006-0190-6}.
\end{barticle}
\endbibitem

\bibitem[\protect\citeauthoryear{{Kosovichev}}{2009}]{Kosovichev2009b}
\begin{barticle}
\bauthor{\bsnm{{Kosovichev}}, \binits{A.G.}}:
\byear{2009},
\bjtitle{Space Science Rev.}
\bvolume{144},
\bfpage{175}.
\url{doi:10.1007/s11214-009-9487-8}.
\end{barticle}
\endbibitem

\bibitem[\protect\citeauthoryear{{Kosovichev} and
  {Duvall}}{2006}]{Kosovichev2006}
\begin{barticle}
\bauthor{\bsnm{{Kosovichev}}, \binits{A.G.}}, \bauthor{\bsnm{{Duvall}},
  \binits{T.L.}}:
\byear{2006},
\bjtitle{Space Science Rev.}
\bvolume{124},
\bfpage{1}.
\url{doi:10.1007/s11214-006-9112-z}.
\end{barticle}
\endbibitem

\bibitem[\protect\citeauthoryear{{Kosovichev} and
  {Duvall}}{1997}]{Kosovichev1997}
\begin{botherref}
\oauthor{\bsnm{{Kosovichev}}, \binits{A.G.}}, \oauthor{\bsnm{{Duvall}},
  \binits{T.L.} \bsuffix{Jr.}}:
1997,
In: {Pijpers~F.~P., Christensen-Dalsgaard~J., and Rosenthal~C.~S.} (eds)
\textit{SCORe'96 : Solar Convection and Oscillations and their Relationship},
\textit{Astrophys. Space Science Lib.}, Kluwer Acad. Publ.,
\textbf{225},
241.
\end{botherref}
\endbibitem

\bibitem[\protect\citeauthoryear{{Kosovichev}, {Duvall}, and
  {Scherrer}}{2000}]{Kosovichev2000}
\begin{barticle}
\bauthor{\bsnm{{Kosovichev}}, \binits{A.G.}}, \bauthor{\bsnm{{Duvall}},
  \binits{T.L..} \bsuffix{Jr.}}, \bauthor{\bsnm{{Scherrer}}, \binits{P.H.}}:
\byear{2000},
\bjtitle{\solphys}
\bvolume{192},
\bfpage{159}.
\end{barticle}
\endbibitem

\bibitem[\protect\citeauthoryear{{Kosovichev}
  \textit{et~al.}}{2009}]{Kosovichev2009c}
\begin{botherref}
\oauthor{\bsnm{{Kosovichev}}, \binits{A.G.}}, \oauthor{\bsnm{{Zhao}},
  \binits{J.}}, \oauthor{\bsnm{{Sekii}}, \binits{T.}},
  \oauthor{\bsnm{{Nagashima}}, \binits{K.}}, \oauthor{\bsnm{{Mitra-Kraev}},
  \binits{U.}}:
2009,
In: {Dikpati~M., Arentoft~T., Gonz{\'a}lez Hern{\'a}ndez,~I, Lindsey,~C. \&
  Hill,~F.} (eds)
\textit{Solar-Stellar Dynamos as Revealed by Helio- and Asteroseismology: GONG 2008/SOHO 21},
Astron. Soc. Pacific, San Francisco,
\textbf{416},
41.
\end{botherref}
\endbibitem

\bibitem[Kosovichev \emph{et al.}(2010)]{Kosovichev2010}Kosovichev,
A.G., Basu, S., Bogart, R., Duvall, T.L., Jr, Gonzalez-Hernandez, I.,
Haber, D., Hartlep, T., Howe, R., Komm, R., Kholikov, S., Parchevsky, K.V.,
Tripathy, S., and Zhao, J.: 2010, {\it ArXiv e-prints}, arXiv:1011.0799,
 \textit{Proc. GONG 2010 - SoHO 24:
A New Era of Seismology of the Sun and Solar-like Stars}, \textit{J. Phys.: Conf. Ser.}, in press.



\bibitem[\protect\citeauthoryear{{Kosugi} \textit{et~al.}}{2007}]{Kosugi2007}
\begin{barticle}
\bauthor{\bsnm{{Kosugi}}, \binits{T.}}, \bauthor{\bsnm{{Matsuzaki}},
  \binits{K.}}, \bauthor{\bsnm{{Sakao}}, \binits{T.}},
  \bauthor{\bsnm{{Shimizu}}, \binits{T.}}, \bauthor{\bsnm{{Sone}},
  \binits{Y.}}, \bauthor{\bsnm{{Tachikawa}}, \binits{S.}},
  \bauthor{\bsnm{{Hashimoto}}, \binits{T.}}, \bauthor{\bsnm{{Minesugi}},
  \binits{K.}}, \bauthor{\bsnm{{Ohnishi}}, \binits{A.}},
  \bauthor{\bsnm{{Yamada}}, \binits{T.}}, \bauthor{\bsnm{{Tsuneta}},
  \binits{S.}}, \bauthor{\bsnm{{Hara}}, \binits{H.}},
  \bauthor{\bsnm{{Ichimoto}}, \binits{K.}}, \bauthor{\bsnm{{Suematsu}},
  \binits{Y.}}, \bauthor{\bsnm{{Shimojo}}, \binits{M.}},
  \bauthor{\bsnm{{Watanabe}}, \binits{T.}}, \bauthor{\bsnm{{Shimada}},
  \binits{S.}}, \bauthor{\bsnm{{Davis}}, \binits{J.M.}},
  \bauthor{\bsnm{{Hill}}, \binits{L.D.}}, \bauthor{\bsnm{{Owens}},
  \binits{J.K.}}, \bauthor{\bsnm{{Title}}, \binits{A.M.}},
  \bauthor{\bsnm{{Culhane}}, \binits{J.L.}}, \bauthor{\bsnm{{Harra}},
  \binits{L.K.}}, \bauthor{\bsnm{{Doschek}}, \binits{G.A.}},
  \bauthor{\bsnm{{Golub}}, \binits{L.}}:
\byear{2007},
\bjtitle{\solphys}
\bvolume{243},
\bfpage{3}.
\url{doi:10.1007/s11207-007-9014-6}.
\end{barticle}
\endbibitem

\bibitem[\protect\citeauthoryear{{Maltby} \textit{et~al.}}{1986}]{Maltby1986}
\begin{barticle}
\bauthor{\bsnm{{Maltby}}, \binits{P.}}, \bauthor{\bsnm{{Avrett}},
  \binits{E.H.}}, \bauthor{\bsnm{{Carlsson}}, \binits{M.}},
  \bauthor{\bsnm{{Kjeldseth-Moe}}, \binits{O.}}, \bauthor{\bsnm{{Kurucz}},
  \binits{R.L.}}, \bauthor{\bsnm{{Loeser}}, \binits{R.}}:
\byear{1986},
\bjtitle{\apj}
\bvolume{306},
\bfpage{284}.
\url{doi:10.1086/164342}.
\end{barticle}
\endbibitem

\bibitem[\protect\citeauthoryear{{Mart{\'{\i}}nez Pillet}
  \textit{et~al.}}{2009}]{MartinezPillet2009}
\begin{barticle}
\bauthor{\bsnm{{Mart{\'{\i}}nez Pillet}}, \binits{V.}},
  \bauthor{\bsnm{{Katsukawa}}, \binits{Y.}}, \bauthor{\bsnm{{Puschmann}},
  \binits{K.G.}}, \bauthor{\bsnm{{Ruiz Cobo}}, \binits{B.}}:
\byear{2009},
\bjtitle{\apjl}
\bvolume{701},
\bfpage{L79}.
\url{doi:10.1088/0004-637X/701/2/L79}.
\end{barticle}
\endbibitem

\bibitem[\protect\citeauthoryear{{Meyer}, {Schmidt}, and
  {Weiss}}{1977}]{Meyer1977}
\begin{barticle}
\bauthor{\bsnm{{Meyer}}, \binits{F.}}, \bauthor{\bsnm{{Schmidt}},
  \binits{H.U.}}, \bauthor{\bsnm{{Weiss}}, \binits{N.O.}}:
\byear{1977},
\bjtitle{\mnras}
\bvolume{179},
\bfpage{741}.
\end{barticle}
\endbibitem

\bibitem[Moradi \emph{et al.}(2010)]{Moradi2010}Moradi, H.,
Baldner, C., Birch, A.C., Braun, D.C., Cameron, R.H., Duvall, T.L., Gizon,
L., Haber, D., Hanasoge, S.M., Hindman, B.W., Jackiewicz, J., Khomenko, E.,
Komm, R., Rajaguru, P., Rempel, M., Roth, M., Schlichenmaier, R., Schunker,
H., Spruit, H.C., Strassmeier, K.G., Thompson, M.J., and Zharkov, S.: 2010,
{\it Solar Phys.} {\bf 267}, 1.

\bibitem[\protect\citeauthoryear{{Moreno-Insertis} and
  {Spruit}}{1989}]{Moreno-Insertis1989}
\begin{barticle}
\bauthor{\bsnm{{Moreno-Insertis}}, \binits{F.}}, \bauthor{\bsnm{{Spruit}},
  \binits{H.C.}}:
\byear{1989},
\bjtitle{\apj}
\bvolume{342},
\bfpage{1158}.
\url{doi:10.1086/167673}.
\end{barticle}
\endbibitem

\bibitem[\protect\citeauthoryear{{Nigam} and {Kosovichev}}{2010}]{Nigam2010}
\begin{barticle}
\bauthor{\bsnm{{Nigam}}, \binits{R.}}, \bauthor{\bsnm{{Kosovichev}},
  \binits{A.G.}}:
\byear{2010},
\bjtitle{\apj}
\bvolume{708},
\bfpage{1475}.
\url{doi:10.1088/0004-637X/708/2/1475}.
\end{barticle}
\endbibitem

\bibitem[\protect\citeauthoryear{{Olshevsky}, {Khomenko}, and
  {Collados}}{2008}]{Olshevsky2008}
\begin{barticle}
\bauthor{\bsnm{{Olshevsky}}, \binits{V.}}, \bauthor{\bsnm{{Khomenko}},
  \binits{E.}}, \bauthor{\bsnm{{Collados}}, \binits{M.}}:
\byear{2008},
\bjtitle{12th European Solar Physics Meeting}, \url{http://espm.kis.uni-freiburg.de/}, p.3.2.
\end{barticle}
\endbibitem

\bibitem[\protect\citeauthoryear{{Parchevsky} and
  {Kosovichev}}{2009}]{Parchevsky2009}
\begin{barticle}
\bauthor{\bsnm{{Parchevsky}}, \binits{K.V.}}, \bauthor{\bsnm{{Kosovichev}},
  \binits{A.G.}}:
\byear{2009},
\bjtitle{\apj}
\bvolume{694},
\bfpage{573}.
\url{doi:10.1088/0004-637X/694/1/573}.
\end{barticle}
\endbibitem

\bibitem[\protect\citeauthoryear{{Parchevsky}, {Zhao}, and
  {Kosovichev}}{2008}]{Parchevsky2008}
\begin{barticle}
\bauthor{\bsnm{{Parchevsky}}, \binits{K.V.}}, \bauthor{\bsnm{{Zhao}},
  \binits{J.}}, \bauthor{\bsnm{{Kosovichev}}, \binits{A.G.}}:
\byear{2008},
\bjtitle{\apj}
\bvolume{678},
\bfpage{1498}.
\url{doi:10.1086/533495}.
\end{barticle}
\endbibitem

\bibitem[\protect\citeauthoryear{{Parchevsky}
  \textit{et~al.}}{2010}]{Parchevsky2010}
\begin{botherref}
\oauthor{\bsnm{{Parchevsky}}, \binits{K.}}, \oauthor{\bsnm{{Kosovichev}},
  \binits{A.}}, \oauthor{\bsnm{{Khomenko}}, \binits{E.}},
  \oauthor{\bsnm{{Olshevsky}}, \binits{V.}}, \oauthor{\bsnm{{Collados}},
  \binits{M.}}:
2010,
\textit{ArXiv e-prints}
\textbf{1002.1117}.
\end{botherref}
\endbibitem

\bibitem[\protect\citeauthoryear{{Parker}}{1979}]{Parker1979}
\begin{barticle}
\bauthor{\bsnm{{Parker}}, \binits{E.N.}}:
\byear{1979},
\bjtitle{\apj}
\bvolume{230},
\bfpage{905}.
\url{doi:10.1086/157150}.
\end{barticle}
\endbibitem

\bibitem[\protect\citeauthoryear{{Ponomarenko}}{1972}]{Ponomarenko1972}
\begin{barticle}
\bauthor{\bsnm{{Ponomarenko}}, \binits{Y.B.}}:
\byear{1972},
\bjtitle{Soviet Astron.}
\bvolume{16},
\bfpage{116}.
\end{barticle}
\endbibitem

\bibitem[\protect\citeauthoryear{{Rajaguru}~\textit{et~al.}}{2006}]{Rajaguru2006}
\begin{barticle}
\bauthor{\bsnm{{Rajaguru}}, \binits{S.P.}}, \bauthor{\bsnm{{Birch}},
  \binits{A.C.}}, \bauthor{\bsnm{{Duvall}}, \binits{T.L.} \bsuffix{Jr.}},
  \bauthor{\bsnm{{Thompson}}, \binits{M.J.}}, \bauthor{\bsnm{{Zhao}},
  \binits{J.}}:
\byear{2006},
\bjtitle{\apj}
\bvolume{646},
\bfpage{543}.
\url{doi:10.1086/504705}.
\end{barticle}
\endbibitem

\bibitem[\protect\citeauthoryear{{Rajaguru}
  \textit{et~al.}}{2007}]{Rajaguru2007}
\begin{barticle}
\bauthor{\bsnm{{Rajaguru}}, \binits{S.P.}},
  \bauthor{\bsnm{{Sankarasubramanian}}, \binits{K.}},
  \bauthor{\bsnm{{Wachter}}, \binits{R.}}, \bauthor{\bsnm{{Scherrer}},
  \binits{P.H.}}:
\byear{2007},
\bjtitle{\apjl}
\bvolume{654},
\bfpage{L175}.
\url{doi:10.1086/511266}.
\end{barticle}
\endbibitem

\bibitem[\protect\citeauthoryear{{Rempel} \textit{et~al.}}{2009}]{Rempel2009}
\begin{barticle}
\bauthor{\bsnm{{Rempel}}, \binits{M.}}, \bauthor{\bsnm{{Sch{\"u}ssler}},
  \binits{M.}}, \bauthor{\bsnm{{Cameron}}, \binits{R.H.}},
  \bauthor{\bsnm{{Kn{\"o}lker}}, \binits{M.}}:
\byear{2009},
\bjtitle{Science}
\bvolume{325},
\bfpage{171}.
\url{doi:10.1126/science.1173798}.
\end{barticle}
\endbibitem

\bibitem[\protect\citeauthoryear{{Sainz Dalda} and {Bellot
  Rubio}}{2008}]{SainzDalda2008}
\begin{barticle}
\bauthor{\bsnm{{Sainz Dalda}}, \binits{A.}}, \bauthor{\bsnm{{Bellot Rubio}},
  \binits{L.R.}}:
\byear{2008},
\bjtitle{\aap}
\bvolume{481},
\bfpage{L21}.
\url{doi:10.1051/0004-6361:20079115}.
\end{barticle}
\endbibitem

\bibitem[\protect\citeauthoryear{{Sainz Dalda} and {Mart{\'{\i}}nez
  Pillet}}{2005}]{SainzDalda2005}
\begin{barticle}
\bauthor{\bsnm{{Sainz Dalda}}, \binits{A.}}, \bauthor{\bsnm{{Mart{\'{\i}}nez
  Pillet}}, \binits{V.}}:
\byear{2005},
\bjtitle{\apj}
\bvolume{632},
\bfpage{1176}.
\url{doi:10.1086/433168}.
\end{barticle}
\endbibitem

\bibitem[\protect\citeauthoryear{{Sankarasubramanian} and
  {Rimmele}}{2003}]{Sankarasubramanian2003}
\begin{barticle}
\bauthor{\bsnm{{Sankarasubramanian}}, \binits{K.}}, \bauthor{\bsnm{{Rimmele}},
  \binits{T.}}:
\byear{2003},
\bjtitle{\apj}
\bvolume{598},
\bfpage{689}.
\url{doi:10.1086/378883}.
\end{barticle}
\endbibitem

\bibitem[\protect\citeauthoryear{{Scharmer}}{2009}]{Scharmer2009}
\begin{barticle}
\bauthor{\bsnm{{Scharmer}}, \binits{G.B.}}:
\byear{2009},
\bjtitle{Space Science Rev.}
\bvolume{144},
\bfpage{229}.
\url{doi:10.1007/s11214-008-9483-4}.
\end{barticle}
\endbibitem

\bibitem[\protect\citeauthoryear{{Scharmer}, {Nordlund}, and
  {Heinemann}}{2008}]{Scharmer2008}
\begin{barticle}
\bauthor{\bsnm{{Scharmer}}, \binits{G.B.}}, \bauthor{\bsnm{{Nordlund}},
  \binits{{\AA}.}}, \bauthor{\bsnm{{Heinemann}}, \binits{T.}}:
\byear{2008},
\bjtitle{\apjl}
\bvolume{677},
\bfpage{L149}.
\url{doi:10.1086/587982}.
\end{barticle}
\endbibitem

\bibitem[\protect\citeauthoryear{{Schmidt}}{1968}]{Schmidt1968}
\begin{botherref}
\oauthor{\bsnm{{Schmidt}}, \binits{H.U.}}:
1968,
In: {Kiepenheuer~K.~O.} (ed.)
\textit{Structure and Development of Solar Active Regions},
\textit{IAU Symposium}, D. Reidel, Dordrecht,
\textbf{35},
95.
\end{botherref}
\endbibitem

\bibitem[\protect\citeauthoryear{{Schunker}, {Braun}, and
  {Cally}}{2007}]{Schunker2007}
\begin{barticle}
\bauthor{\bsnm{{Schunker}}, \binits{H.}}, \bauthor{\bsnm{{Braun}},
  \binits{D.C.}}, \bauthor{\bsnm{{Cally}}, \binits{P.S.}}:
\byear{2007},
\bjtitle{Astron. Nach.}
\bvolume{328},
\bfpage{292}.
\url{doi:10.1002/asna.200610732}.
\end{barticle}
\endbibitem

\bibitem[\protect\citeauthoryear{{Schunker}
  \textit{et~al.}}{2005}]{Schunker2005}
\begin{barticle}
\bauthor{\bsnm{{Schunker}}, \binits{H.}}, \bauthor{\bsnm{{Braun}},
  \binits{D.C.}}, \bauthor{\bsnm{{Cally}}, \binits{P.S.}},
  \bauthor{\bsnm{{Lindsey}}, \binits{C.}}:
\byear{2005},
\bjtitle{\apjl}
\bvolume{621},
\bfpage{L149}.
\url{doi:10.1086/429290}.
\end{barticle}
\endbibitem

\bibitem[\protect\citeauthoryear{{Sheeley}}{1969}]{Sheeley1969}
\begin{barticle}
\bauthor{\bsnm{{Sheeley}}, \binits{N.R.} \bsuffix{Jr.}}:
\byear{1969},
\bjtitle{\solphys}
\bvolume{9},
\bfpage{347}.
\url{doi:10.1007/BF02391657}.
\end{barticle}
\endbibitem

\bibitem[\protect\citeauthoryear{{Sheeley}}{1972}]{Sheeley1972}
\begin{barticle}
\bauthor{\bsnm{{Sheeley}}, \binits{N.R.} \bsuffix{Jr.}}:
\byear{1972},
\bjtitle{\solphys}
\bvolume{25},
\bfpage{98}.
\url{doi:10.1007/BF00155747}.
\end{barticle}
\endbibitem

\bibitem[\protect\citeauthoryear{{Shelyag}
  \textit{et~al.}}{2009{\natexlab{a}}}]{Shelyag2009}
\begin{barticle}
\bauthor{\bsnm{{Shelyag}}, \binits{S.}}, \bauthor{\bsnm{{Zharkov}},
  \binits{S.}}, \bauthor{\bsnm{{Fedun}}, \binits{V.}},
  \bauthor{\bsnm{{Erd{\'e}lyi}}, \binits{R.}}, \bauthor{\bsnm{{Thompson}},
  \binits{M.J.}}:
\byear{2009},
\bjtitle{\aap}
\bvolume{501},
\bfpage{735}.
\url{doi:10.1051/0004-6361/200911709}.
\end{barticle}
\endbibitem

\bibitem[\protect\citeauthoryear{{Shelyag}
  \textit{et~al.}}{2009{\natexlab{b}}}]{Shelyag2009a}
\begin{botherref}
\oauthor{\bsnm{{Shelyag}}, \binits{S.}}, \oauthor{\bsnm{{Zharkov}},
  \binits{S.}}, \oauthor{\bsnm{{Fedun}}, \binits{V.}},
  \oauthor{\bsnm{{Erd{\'e}lyi}}, \binits{R.}}, \oauthor{\bsnm{{Thompson}},
  \binits{M.J.}}:
2009,
In: {Dikpati~M., Arentoft~T., Gonz{\'a}lez Hern{\'a}ndez,~I, Lindsey,~C. \&
  Hill,~F.} (eds)
\textit{Solar-Stellar Dynamos as Revealed by Helio- and Asteroseismology: GONG 2008/SOHO 21},
Astron. Soc. Pacific, San Francisco,
\textbf{416},
167.
\end{botherref}
\endbibitem

\bibitem[\protect\citeauthoryear{{Sobotka} and {Roudier}}{2007}]{Sobotka2007}
\begin{barticle}
\bauthor{\bsnm{{Sobotka}}, \binits{M.}}, \bauthor{\bsnm{{Roudier}},
  \binits{T.}}:
\byear{2007},
\bjtitle{\aap}
\bvolume{472},
\bfpage{277}.
\url{doi:10.1051/0004-6361:20077552}.
\end{barticle}
\endbibitem

\bibitem[\protect\citeauthoryear{{Stein}, {Bercik}, and
  {Nordlund}}{2003}]{Stein2003}
\begin{botherref}
\oauthor{\bsnm{{Stein}}, \binits{R.F.}}, \oauthor{\bsnm{{Bercik}},
  \binits{D.}}, \oauthor{\bsnm{{Nordlund}}, \binits{{\AA}.}}:
2003,
In: {Pevtsov~A.~A. and Uitenbroek~H.} (ed.)
\textit{Current Theoretical Models and Future High Resolution Solar
  Observations: Preparing for ATST},
\textrm{Astron. Soc. Pacific, San Francisco}
\textbf{286},
121.
\end{botherref}
\endbibitem

\bibitem[\protect\citeauthoryear{{Sun}, {Chou}, and {TON
  Team}}{2002}]{Sun2002}
\begin{barticle}
\bauthor{\bsnm{{Sun}}, \binits{M.}}, \bauthor{\bsnm{{Chou}}, \binits{D.}},
  \bauthor{\bsnm{{The TON Team}}}:
\byear{2002},
\bjtitle{\solphys}
\bvolume{209},
\bfpage{5}.
\url{doi:10.1023/A:1020909524039}.
\end{barticle}
\endbibitem

\bibitem[\protect\citeauthoryear{{Tsuneta} \textit{et~al.}}{2008}]{Tsuneta2008}
\begin{barticle}
\bauthor{\bsnm{{Tsuneta}}, \binits{S.}}, \bauthor{\bsnm{{Ichimoto}},
  \binits{K.}}, \bauthor{\bsnm{{Katsukawa}}, \binits{Y.}},
  \bauthor{\bsnm{{Nagata}}, \binits{S.}}, \bauthor{\bsnm{{Otsubo}},
  \binits{M.}}, \bauthor{\bsnm{{Shimizu}}, \binits{T.}},
  \bauthor{\bsnm{{Suematsu}}, \binits{Y.}}, \bauthor{\bsnm{{Nakagiri}},
  \binits{M.}}, \bauthor{\bsnm{{Noguchi}}, \binits{M.}},
  \bauthor{\bsnm{{Tarbell}}, \binits{T.}}, \bauthor{\bsnm{{Title}},
  \binits{A.}}, \bauthor{\bsnm{{Shine}}, \binits{R.}},
  \bauthor{\bsnm{{Rosenberg}}, \binits{W.}}, \bauthor{\bsnm{{Hoffmann}},
  \binits{C.}}, \bauthor{\bsnm{{Jurcevich}}, \binits{B.}},
  \bauthor{\bsnm{{Kushner}}, \binits{G.}}, \bauthor{\bsnm{{Levay}},
  \binits{M.}}, \bauthor{\bsnm{{Lites}}, \binits{B.}},
  \bauthor{\bsnm{{Elmore}}, \binits{D.}}, \bauthor{\bsnm{{Matsushita}},
  \binits{T.}}, \bauthor{\bsnm{{Kawaguchi}}, \binits{N.}},
  \bauthor{\bsnm{{Saito}}, \binits{H.}}, \bauthor{\bsnm{{Mikami}},
  \binits{I.}}, \bauthor{\bsnm{{Hill}}, \binits{L.D.}},
  \bauthor{\bsnm{{Owens}}, \binits{J.K.}}:
\byear{2008},
\bjtitle{\solphys}
\bvolume{249},
\bfpage{167}.
\url{doi:10.1007/s11207-008-9174-z}.
\end{barticle}
\endbibitem

\bibitem[\protect\citeauthoryear{{Vargas Dom{\'{\i}}nguez}
  \textit{et~al.}}{2007}]{VargasDominguez2007}
\begin{barticle}
\bauthor{\bsnm{{Vargas Dom{\'{\i}}nguez}}, \binits{S.}},
  \bauthor{\bsnm{{Bonet}}, \binits{J.A.}}, \bauthor{\bsnm{{Mart{\'{\i}}nez
  Pillet}}, \binits{V.}}, \bauthor{\bsnm{{Katsukawa}}, \binits{Y.}},
  \bauthor{\bsnm{{Kitakoshi}}, \binits{Y.}}, \bauthor{\bsnm{{Rouppe van der
  Voort}}, \binits{L.}}:
\byear{2007},
\bjtitle{\apjl}
\bvolume{660},
\bfpage{L165}.
\url{doi:10.1086/518123}.
\end{barticle}
\endbibitem

\bibitem[\protect\citeauthoryear{{Vargas Dom{\'{\i}}nguez}
  \textit{et~al.}}{2008}]{VargasDominguez2008}
\begin{barticle}
\bauthor{\bsnm{{Vargas Dom{\'{\i}}nguez}}, \binits{S.}}, \bauthor{\bsnm{{Rouppe
  van der Voort}}, \binits{L.}}, \bauthor{\bsnm{{Bonet}}, \binits{J.A.}},
  \bauthor{\bsnm{{Mart{\'{\i}}nez Pillet}}, \binits{V.}}, \bauthor{\bsnm{{Van
  Noort}}, \binits{M.}}, \bauthor{\bsnm{{Katsukawa}}, \binits{Y.}}:
\byear{2008},
\bjtitle{\apj}
\bvolume{679},
\bfpage{900}.
\url{doi:10.1086/587139}.
\end{barticle}
\endbibitem

\bibitem[\protect\citeauthoryear{{Vargas Dominguez}
  \textit{et~al.}}{2010}]{VargasDominguez2010}
\begin{botherref}
\oauthor{\bsnm{{Vargas Dominguez}}, \binits{S.}}, \oauthor{\bsnm{{de Vicente}},
  \binits{A.}}, \oauthor{\bsnm{{Bonet}}, \binits{J.A.}},
  \oauthor{\bsnm{{Martinez Pillet}}, \binits{V.}}:
\byear{2010},
\bjtitle{\aa}
\bvolume{516},
\bfpage{A91}.
\end{botherref}
\endbibitem

\bibitem[\protect\citeauthoryear{{V{\"o}gler}
  \textit{et~al.}}{2005}]{Vogler2005}
\begin{barticle}
\bauthor{\bsnm{{V{\"o}gler}}, \binits{A.}}, \bauthor{\bsnm{{Shelyag}},
  \binits{S.}}, \bauthor{\bsnm{{Sch{\"u}ssler}}, \binits{M.}},
  \bauthor{\bsnm{{Cattaneo}}, \binits{F.}}, \bauthor{\bsnm{{Emonet}},
  \binits{T.}}, \bauthor{\bsnm{{Linde}}, \binits{T.}}:
\byear{2005},
\bjtitle{\aap}
\bvolume{429},
\bfpage{335}.
\url{doi:10.1051/0004-6361:20041507}.
\end{barticle}
\endbibitem

\bibitem[\protect\citeauthoryear{{Wachter}, {Rajaguru}, and
  {Bogart}}{2006{\natexlab{a}}}]{Wachter2006a}
\begin{botherref}
\oauthor{\bsnm{{Wachter}}, \binits{R.}}, \oauthor{\bsnm{{Rajaguru}},
  \binits{S.P.}}, \oauthor{\bsnm{{Bogart}}, \binits{R.S.}}:
2006,
In: \textit{Proceedings of SOHO 18/GONG 2006/HELAS I, Beyond the spherical
  Sun},
SP-\textbf{624}, ESA, Noordwijk, p.48.1.
\end{botherref}
\endbibitem

\bibitem[\protect\citeauthoryear{{Wachter}, {Schou}, and
  {Sankarasubramanian}}{2006{\natexlab{b}}}]{Wachter2006}
\begin{barticle}
\bauthor{\bsnm{{Wachter}}, \binits{R.}}, \bauthor{\bsnm{{Schou}}, \binits{J.}},
  \bauthor{\bsnm{{Sankarasubramanian}}, \binits{K.}}:
\byear{2006},
\bjtitle{\apj}
\bvolume{648},
\bfpage{1256}.
\url{doi:10.1086/505930}.
\end{barticle}
\endbibitem

\bibitem[\protect\citeauthoryear{{Zhao} and
  {Kosovichev}}{2003{\natexlab{a}}}]{Zhao2003}
\begin{barticle}
\bauthor{\bsnm{{Zhao}}, \binits{J.}}, \bauthor{\bsnm{{Kosovichev}},
  \binits{A.G.}}:
\byear{2003},
\bjtitle{\apj}
\bvolume{591},
\bfpage{446}.
\url{doi:10.1086/375343}.
\end{barticle}
\endbibitem

\bibitem[\protect\citeauthoryear{{Zhao} and
  {Kosovichev}}{2003{\natexlab{b}}}]{Zhao2003a}
\begin{botherref}
\oauthor{\bsnm{{Zhao}}, \binits{J.}}, \oauthor{\bsnm{{Kosovichev}},
  \binits{A.G.}}:
2003,
In: {Sawaya-Lacoste, ~H.} (ed.)
\textit{GONG+ 2002. Local and Global Helioseismology: the Present and Future},
SP-\textbf{517}, ESA, Noordwijk,
417
\end{botherref}
\endbibitem

\bibitem[\protect\citeauthoryear{{Zhao} and {Kosovichev}}{2006}]{Zhao2006}
\begin{barticle}
\bauthor{\bsnm{{Zhao}}, \binits{J.}}, \bauthor{\bsnm{{Kosovichev}},
  \binits{A.G.}}:
\byear{2006},
\bjtitle{\apj}
\bvolume{643},
\bfpage{1317}.
\url{doi:10.1086/503248}.
\end{barticle}
\endbibitem

\bibitem[\protect\citeauthoryear{{Zhao}, {Kosovichev}, and
  {Duvall}}{2001}]{Zhao2001}
\begin{barticle}
\bauthor{\bsnm{{Zhao}}, \binits{J.}}, \bauthor{\bsnm{{Kosovichev}},
  \binits{A.G.}}, \bauthor{\bsnm{{Duvall}}, \binits{T.L.} \bsuffix{Jr.}}:
\byear{2001},
\bjtitle{\apj}
\bvolume{557},
\bfpage{384}.
\url{doi:10.1086/321491}.
\end{barticle}
\endbibitem

\bibitem[\protect\citeauthoryear{{Zhao}, {Kosovichev}, and
  {Ilonidis}}{2010{\natexlab{a}}}]{Zhao2010a}
\begin{botherref}
\oauthor{\bsnm{{Zhao}}, \binits{J.}}, \oauthor{\bsnm{{Kosovichev}},
  \binits{A.G.}}, \oauthor{\bsnm{{Ilonidis}}, \binits{S.}}:
2010,
\textit{\solphys}
\textrm{in press}.
\end{botherref}
\endbibitem

\bibitem[\protect\citeauthoryear{{Zhao}, {Kosovichev}, and
  {Sekii}}{2010{\natexlab{b}}}]{Zhao2010}
\begin{barticle}
\bauthor{\bsnm{{Zhao}}, \binits{J.}}, \bauthor{\bsnm{{Kosovichev}},
  \binits{A.G.}}, \bauthor{\bsnm{{Sekii}}, \binits{T.}}:
\byear{2010},
\bjtitle{\apj}
\bvolume{708},
\bfpage{304}.
\url{doi:10.1088/0004-637X/708/1/304}.
\end{barticle}
\endbibitem

\bibitem[\protect\citeauthoryear{{Zharkov}, {Nicholas}, and
  {Thompson}}{2007}]{Zharkov2007}
\begin{barticle}
\bauthor{\bsnm{{Zharkov}}, \binits{S.}}, \bauthor{\bsnm{{Nicholas}},
  \binits{C.J.}}, \bauthor{\bsnm{{Thompson}}, \binits{M.J.}}:
\byear{2007},
\bjtitle{Astron. Nach.}
\bvolume{328},
\bfpage{240}.
\url{doi:10.1002/asna.200610744}.
\end{barticle}
\endbibitem

\bibitem[\protect\citeauthoryear{{Zuccarello}
  \textit{et~al.}}{2009}]{Zuccarello2009}
\begin{barticle}
\bauthor{\bsnm{{Zuccarello}}, \binits{F.}}, \bauthor{\bsnm{{Romano}},
  \binits{P.}}, \bauthor{\bsnm{{Guglielmino}}, \binits{S.L.}},
  \bauthor{\bsnm{{Centrone}}, \binits{M.}}, \bauthor{\bsnm{{Criscuoli}},
  \binits{S.}}, \bauthor{\bsnm{{Ermolli}}, \binits{I.}},
  \bauthor{\bsnm{{Berrilli}}, \binits{F.}}, \bauthor{\bsnm{{Del Moro}},
  \binits{D.}}:
\byear{2009},
\bjtitle{\aap}
\bvolume{500},
\bfpage{L5}.
\url{doi:10.1051/0004-6361/200912277}.
\end{barticle}
\endbibitem

\end{thebibliography}

\end{article}
\end{document}